\documentclass{iopart}

\usepackage{lscape}
\usepackage{setspace}
\usepackage{multicol}
\usepackage{enumerate}
\usepackage{subfig}
\usepackage{paralist}
\usepackage{graphicx}
\usepackage{calc,xcolor}
\usepackage{mathptmx,bm}
\usepackage[breaklinks=true]{hyperref}
\usepackage{iopams}

\newcommand{\EMILIA}[1]{\textcolor{cyan}{\fbox{Emilia} {\sl#1}}}

\begin{document}

%\topical{tight binding model}
\title[A tight binding model for MoS$_2$ monolayers]{A tight-binding
  model for MoS$_{2}$ monolayers}

\author{E Ridolfi$^1$, D Le$^2$, T S Rahman$^2$, E R Mucciolo$^2$ and C H
  Lewenkopf$^1$}

\address{$^1$ Instituto de F\a'{i}sica, Universidade Federal
  Fluminense, 24210-346 Niter\a'{o}i, RJ, Brazil}

\address{$^2$ Department of Physics, University of Central Florida,
  Orlando, FL 32816-2385, USA}

\ead{\mailto{emilia.ridolfi@gmail.com} and
  \mailto{mucciolo@physics.ucf.edu} and \mailto{caio@if.uff.br}}

%%%%%%%%%%%%%%%%%%%%%%%%%%%%%%%%%%%%%%%%%%%%%%%
\begin{abstract}
  We propose an accurate tight-binding parametrization for the band
  structure of MoS$_2$ monolayers near the main energy gap. We
  introduce a generic and straightforward derivation for the band
  energies equations that could be employed for other monolayer
  dichalcogenides. A parametrization that includes spin-orbit coupling
  is also provided. The proposed set of model parameters reproduce
  both the correct orbital compositions and location of valence and
  conductance band in comparison with {\it ab initio}
  calculations. The model gives a suitable starting point for
  realistic large-scale atomistic electronic transport calculations.
\end{abstract}
%%%%%%%%%%%%%%%%%%%%%%%%%%%%%%%%%%%%%%%%%%%%%%%

%\pacs{72.80.Vp,81.05.ue,72.10.Fk,73.22.Pr}

%\submitto{JCMP, Topical Review}

%\maketitle

%%%%%%%%%%%%%%%%%%%%%%%%%%%%%%%%%%%%%%%%%%%%%%%
%                                                      Introduction
\section{Introduction}
\label{sec:1introduction}
%%%%%%%%%%%%%%%%%%%%%%%%%%%%%%%%%%%%%%%%%%%%%%%

The synthesis of graphene in 2004 \cite{Novoselov04,Novoselov05}, the
first single-atom thick material, has boosted the research in
atomically thin two-dimensional (2D) materials. The ability to
manipulate isolated single atomic layers and reassemble them to form
heterostructures layer-by-layer in a precise sequence, opens enormous
possibilities for applications \cite{Geim2013,
  WangKis2012,Butler2Dmaterials,Yazyev2015}. Along this approach,
semiconducting dichalcogenides are promising compounds since they can
be easily exfoliated and present a suitable small gap both in bulk and
as a single layer. In this category of 2D dichalcogenides systems,
monolayer molybdenum disulfide (MoS$_2$) has recently gained attention
for combining an electron mobility comparable to that of graphene
devices with a finite energy gap \cite{Radisavljevic2011}. Unlike its
bulk form, which is an indirect gap semiconductor, monolayer MoS$_2$
has a direct gap \cite{Geim2013,Mak2010}, making it very interesting
for optoelectronics. Another interesting feature is that the
electronic properties appear to be highly sensitive to external
pressure \cite{Nayak2012}, strain \cite{Yue2012,Pena2015}, and
temperature \cite{Radisavljevic2013}, which affect the gap and, under
certain conditions, can also induce a insulator/metal transition. In
addition, the lack of lattice inversion symmetry together with
spin-orbit coupling (SOC) leads to coupled spin and valley physics in
monolayers of MoS$_2$ and other group-VI dichalcogenides
\cite{Xiao2012,Mak2014}, making it possible to control spin and valley
in these materials \cite{Butler2Dmaterials,Hawrylak2012}. Due to their
peculiar band structure, a variety of nanoelectronics applications
\cite{WangKis2012, Butler2Dmaterials} including valleytronics,
spintronics, optoelectronics, and room temperature transistor devices
\cite{Radisavljevic2011} have been suggested for monolayers of
MoS$_2$.

In light of the growing interest in this material, an accurate and yet
reasonably simple model describing the band structure and electronic
properties of MoS$_2$ is highly desirable. So far the electronic
properties of single-layer and few-layer dichalcogenides have been
mainly investigated by means of {\it ab initio} calculations, based on
Density Functional Theory (DFT) \cite{Hawrylak2012,Yue2012}. Such
methods provide valuable information about electronic properties of
pristine dichalcogenide crystals, but are computationally prohibitive
to treat disordered systems with a large number of atoms. To address
the latter, one needs to resort to a simple effective model, such as
the $kp$ Hamiltonian or the tight-biding approximation. In this paper
we choose the latter route, which provides a more accurate description
for the entire band structure than the $kp$ method. Moreover, the
tight-binding model applied to a single-layer MoS$_{2}$ as well as to
similar transition metal dichalcogenides, constitutes a key tool for
further studies of the low-energy electronic transport properties of
these materials, such as the description of the conductivity in
diffusive samples, as well as for evaluating the conductance of
ballistic samples as a function of carrier concentration.

In recent years, a variety of tight-binding models have been proposed
for MoS$_{2}$ monolayers \cite{Zahid2013,Liu2013,Rostami2013}.
Unfortunately, they are neither practical nor sufficiently accurate
for transport calculations. For that purpose, one needs a
tight-binding model with a manageable number of parameters and
interactions that accurately reproduces the {\it ab initio} electronic
properties of the conduction band (CB) near its maximum points and the
the valence band (VB) near its minimal points. Before we present our
results, let us now briefly review the main features of the
tight-binding models for dichalcogenides found so far in the
literature.

An ``all orbital model'' was put forward by Zahid and collaborators
\cite{Zahid2013}. The model includes non-orthogonal $sp^{3}d^{5}$
orbitals, considers only nearest-neighbour hopping matrix elements,
and includes spin orbit coupling. The model has 96 fitting
parameters. The optimization of the Slater-Koster energies
\cite{Slater-Koster} and overlap integrals used in the model are
obtained by a fit to the DFT target band structure. The model shows
good agreement with band structure calculations using the HSE06
functional \cite{Heyd2003,erratumHeyd2003}, but its computational cost
and complexity make it impractical for studying disorder and
electronic transport at large scales.

In contrast, Liu and collaborators \cite{Liu2013} proposed a
three-orbital tight-binding model. The authors consider a
superposition of orbitals $d_{z^{2}}, d_{xy}$, and $d_{x^{2}-y^{2}}$
as orthogonal basis, targeting the main orbital composition around the
$K$ point, which corresponds to the direct gap. Thus, the agreement
between their first nearest-neighbour tight-binding model and the DFT
predictions using both the local density approximation (LDA)
\cite{Ceperley1980,Perdew1981} and the generalized-gradient
approximation (GGA) \cite{Perdew1996} is limited to features in the
vicinity of the $K$ point, missing the local band minimal at the $Q$
point. By including up to the third-nearest neighbour hopping
involving Mo-Mo terms, the agreement with the DFT-GGA band structure
improves substantially. This is achieved at the expense of increasing
the complexity of the model, as the number of fitting parameters goes
from 8 to 19. In transport calculations, the inclusion of higher
neighbour hopping terms implies in an increase in the size of the unit
cell. Hence, trading a larger number of bands with nearest-neighbour
hopping for a simpler model with longer range hopping is not
necessarily advantageous. Moreover, the orbital composition in
Ref.~\cite{Liu2013} is, by construction, restricted to Mo orbitals
which limits the analysis of disorder effects. It is also worth noting
that Ref.~\cite{Liu2013} fails to reproduce the orbital composition
and energy spectrum around the $\Gamma$ point, which plays a
significant role in transport for hole-doped monolayers.

A seven-orbitals tight-binding parametrization has been introduced by
Rostami and collaborators \cite{Rostami2013}. The model considers a
non-orthogonal basis and neglects the $s$ and $p_{z}$ orbitals of the
S atoms and the $s$, $d_{yz}$, and $d_{xz}$ orbitals of the Mo atom by
invoking arguments based on crystal symmetry and the range of energies
of interest. The model reproduces the main features around the $K$
point, but two unrealistic flat bands appear in the gap region. We
attribute this undesired feature to the fact that the $p_{z}$ orbital
of the S atoms is not actually decoupled from other orbitals and can
not be neglected. On the contrary, the $p_z$ orbital from S atoms
plays a pivotal role in the transition from a direct to an indirect
gap, when passing from a monolayer to a multilayer system. The basis
set introduced in Ref.~\cite{Rostami2013} does not distinguish between
$p_{z}^{\rm t}$ and $p_{z}^{\rm b}$ $S$ orbitals at the top and bottom
planes of the S--Mo--S layers. Therefore, it can not correctly capture
the symmetry under inversion of the $z$-axis.
 
As pointed out by Cappelluti and collaborators \cite{Cap2013}, the
linear combination of $p_{z}^{\rm t}$ and $p_{z}^{\rm b}$ orbitals is
necessary to produce $z$-symmetric and $z$-antisymmetric states. For
this reason, Refs.~\cite{Cap2013,Cap2014} propose a minimal model of
11 orbitals. This is also our choice. This model considers an
orthogonal basis composed of all the $4d$ Mo orbitals and the $3p$ S
orbitals, forming real symmetric (even) and antisymmetric (odd)
combinations of $p_{x,y,z}$ orbitals \cite{Bromley1972,Mattheis73}.
The tight-binding parameters found in Refs.~\cite{Cap2013,Cap2014}
yield two bands that look very similar to the conduction and the
valence bands obtained by standard DFT calculations. However, in our
treatment, by using analytical expressions for the valence and
conductance bands at high-symmetry $k$-points, we observe that the
tight-binding orbital compositions of Refs. \cite{Cap2013,Cap2014}
have actually no relation with those calculated using DFT. Hence, a
new tight-binding parametrization, reproducing both energies {\it and}
orbital composition is badly needed. This is the main goal of this
paper. We rederive the tight-binding equations of Ref.~\cite{Cap2013}
in a more direct and transparent way, allowing us to more carefully
consider the orbital composition in our parameter optimization
procedure.

The paper is organized as follows. In Sec.~\ref{sec:2structure} we
describe the atomic structure of a MoS$_2$ monolayer and discuss the
DFT-HSE06 results for the the band structure that will be the
reference for our tight-binding model. In Sec.~\ref{sec:3model} we
present the model. In Sec.~\ref{sec:4optimization} we analyze the band
equations for a few high-symmetry $k$-point, allowing us to obtain
simple analytical expressions for the bands. These are used to find
the best set of tight-binding parameters that fit the DFT band
structure. In Sec.~\ref{sec:5-11band} we present the optimized
parameters and the corresponding band structure. In
Sec.~\ref{sec:6easier} we consider a simplified model with a reduced
number of parameters. In Sec.~\ref{sec:7spin} we add spin-orbit
interaction to the full model. Finally, in Sec.~\ref{sec:8conclusion}
we draw our conclusions.

The main text is supplemented by a number of appendices containing
technical aspects of the calculations. In ~\ref{subsec:Appendix A} we
present in detail all the elements required to construct the
tight-binding band equations. In ~\ref{subsec:Appendix B} we show how
to implement the band equations in the unsymmetrized and symmetrized
ones. \ref{subsec:Appendix C} analyzes the band structure at symmetry
points used in the optimization. Finally, in \ref{subsec:Appendix D}
we present a comparison between our 11-band tight-binding formulation
and that of Refs. \cite{Cap2013,Cap2014}.

%%%%%%%%%%%%%%%%%%%%%%%%%%%%%%%%%%%%%%%%%
%   2. Crystal structure and DFT calculation
%
\section{Crystal structure and ab-initio electronic structure}
\label{sec:2structure}
%%%%%%%%%%%%%%%%%%%%%%%%%%%%%%%%%%%%%%%%%

Molybdenum disulfide is a layered transition metal dichalcogenide
semiconductor. The layered structure is formed by a honeycomb
arrangement of Mo and S atoms stacked together and forming S--Mo--S
sandwiches coordinated in a triangular prismatic fashion. The S--Mo--S
layers are bonded together by weak van der Waals forces.

\begin{figure}[h]
\centering
\includegraphics[width=0.5\columnwidth]{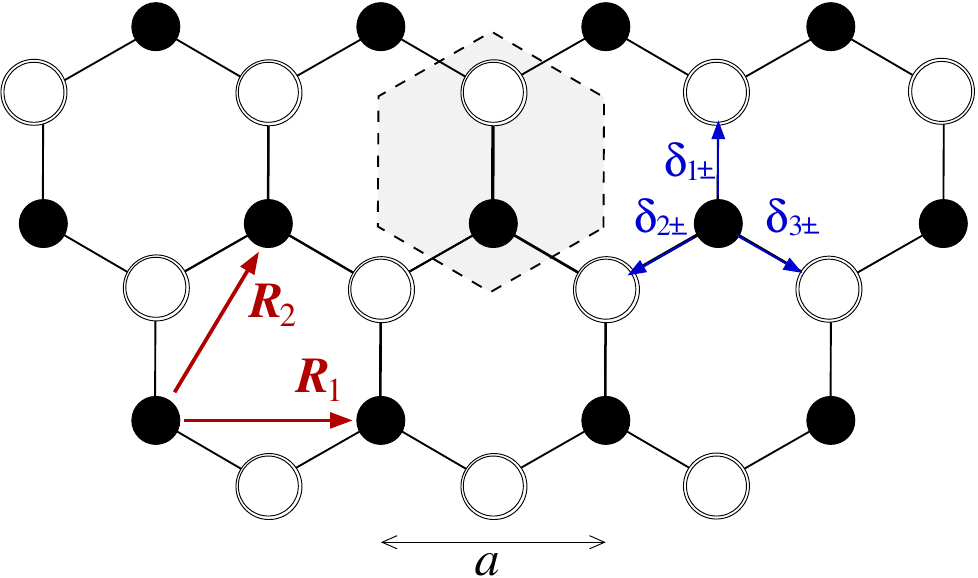}
\hskip0.4cm
\includegraphics[width=0.25\columnwidth]{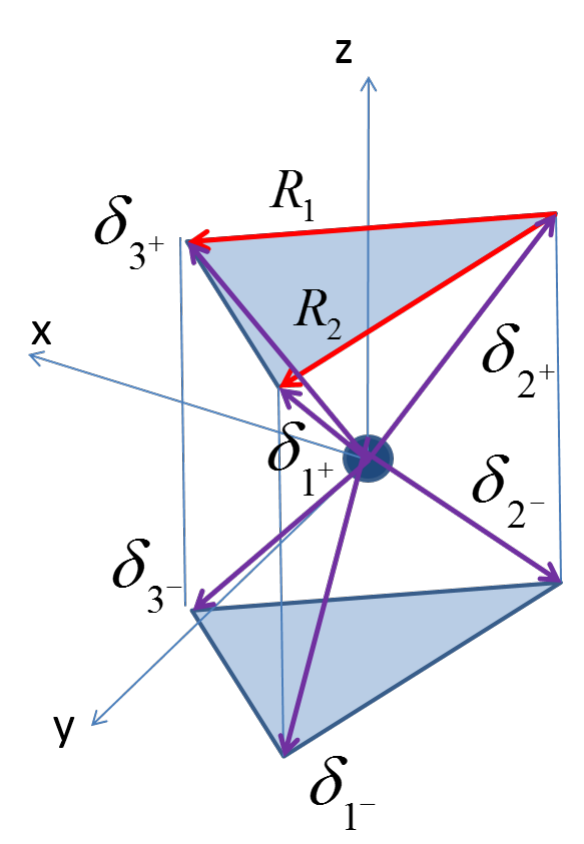}
\caption{Left panel: Top view of the MoS$_{2}$ lattice structure. Dark
  (light) circles represent Mo (S) atoms. Notice that in this view two
  S atoms sit on top of each other. The unit cell is shown in the
  highlighted hexagon. The lattice constant in the Mo plane is
  $a$. The two Bravais lattice vectors (${\bf R}_{1}$ and ${\bf
    R}_{2}$) are indicated. Six other auxiliary vectors that connect a
  Mo atom with its nearest S atoms, ${\bm \delta}_{1\pm}$,
  ${\bm\delta}_{2\pm}$, and ${\bm \delta}_{3\pm}$, are
  indicated. Right panel: Tridimensional view of the first neighbors
  of a Mo atom. The reference trigonal prism coordination
  unit and other useful quantities are also shown.
\label{fig:Top-view-of reticolo}
}
\end{figure}

The single-layer MoS$_{2}$ lattice structure is shown in
Fig.~\ref{fig:Top-view-of reticolo}, top and lateral views. It is a 2D
rhombic lattice with a three-atom basis (one Mo and two S). The two
Bravais primitive lattice vectors are
\begin{equation}
{\bf R}_{1} = \left( a,0,0 \right)
\end{equation}
and
\begin{equation}
%% \quad \mbox{and} \quad 
{\bf R}_{2} = \left( \frac{a}{2},\frac{\sqrt{3}}{2}a,0 \right) ,
\end{equation}
where $a=3.16$ \AA\ is the lattice constant. The S atoms are located
in planes $1.56$ \AA\ above and below the Mo plane. This yields a
distance between neighboring Mo and S atoms of $d = 2.40$ {\AA}. The
angle between the Mo-S bond and the Mo plane is $\theta_{b} =
40.6^{o}$. These values are obtained by the DFT calculation discussed
below and are consistent with previous DFT calculations
\cite{Zahid2013, Yue2012, Hawrylak2012} and with experimental values
\cite{Cap2013, Bromley1972}.

\noappendix 

%-------------------------------------------------------------------------------
%-------------------------------------------------------------------------------
 
\begin{figure}[h]
\centering 
\includegraphics[width=0.35\columnwidth]{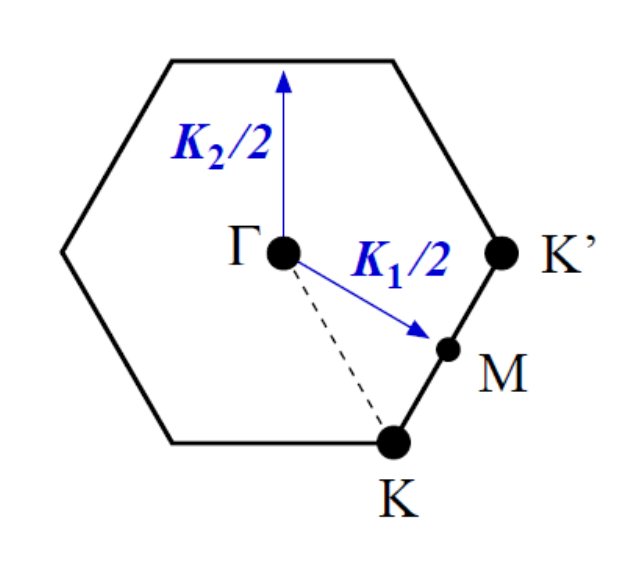}
\caption{Brillouin zone for the MoS$_{2}$ lattice. ${\bf K}_1$ and
  ${\bf K}_2$ are the reciprocal lattice basis vectors, and $\Gamma$,
  $K$, $K'$, and $M$ are the high-symmetry points considered in this
  study.}
\label{fig:Brillouin-zone-for}
\end{figure}

For the purpose of building the tight-binding model, we will follow
the notation introduced Fig.~\ref{fig:Top-view-of reticolo}. We denote
by ``$t$'' (or ``$+$") and by ``$b$" (or ``$-$") the S atoms at the
top and bottom layers, respectively. The distance between the two S
layers is $d\cos\theta_{B}=a/\sqrt{3}$. The nearest-neighbour vectors,
connecting Mo and S atoms, are given by
\begin{eqnarray}
  \bm{\delta}_{1\pm} & = & d \left( 0,\cos\theta_{B},\pm\sin\theta_{B}
  \right), \\ \bm{\delta}_{2\pm} & = & d \left(
    -\frac{\sqrt{3}}{2}\cos\theta_{B}, - \frac{1}{2}\cos\theta_{B},
    \pm\sin\theta_{B} \right), \\ \bm{\delta}_{3\pm} & = & d\left(
    +\frac{\sqrt{3}}{2}\cos\theta_{B}, - \frac{1}{2} 
    \cos\theta_{B},\pm\sin\theta_{B}
  \right).
\end{eqnarray}
The MoS$_{2}$ Brillouin zone is hexagonal. The most important symmetry
points and symmetry lines are indicated in
Fig. \ref{fig:Brillouin-zone-for}, namely, $\Gamma=(0,0)$, $K = \left(
  \frac{2\pi}{3a},\frac{-2\pi}{\sqrt{3}a} \right)$, and $M = \left(
  \frac{\pi}{a},\frac{-\pi}{\sqrt{3}a} \right)$. The reciprocal
lattice basis vectors are
\begin{equation}
{\bf K}_{1} = \frac{4\pi}{\sqrt{3}a} \left(
\frac{\sqrt{3}}{2},-\frac{1}{2},0 \right)
\end{equation}
and
%%\quad \mbox{and} \quad
%
\begin{equation}
{\bf K}_{2} = \frac{4\pi}{\sqrt{3}a} \left( 0,1,0 \right)
\end{equation}

%FIGURE 2
%----------------------------------------------------------------------------
%----------------------------------------------------------------------------

%Duy modified this part with a new table + references
% Thank you for 159 ->149 correction. I spoted that as well. 
%---------- NEW TABLE -----------------------------------------
\begin{table}[h]
\centering
\caption{Summary of experimental and theoretical values of the band gap 
of MoS$_2$.}
\label{tab:band_gap} 
\begin{tabular}{cccc}
\br
Method     & Value (eV)    & Note  \\
\mr
Photoluminescence  \cite{Mak2010,Splendiani2010} &1.8-1.9 &  Optical band gap \\
Scanning Tunneling Spectrocopy (STS)  \cite{Zhang_exciton2014} & 2.15  & \\
DFT-LDA \cite{Torres2015} & 1.81  &    \\
DFT-PBE \cite{Mann2014} & 1.68  &    \\
DFT-optB88-vdW \cite{Torres2015} & 1.67   &   \\
GW \cite{Qiu2013} & 2.84   & G$_1$W$_0$ approximation  \\
DFT-HSE06 & 2.23  &   This work &   \\
\br
\end{tabular}
\end{table}
%----------- end TABLE 1 [IOP format] ----------------------------------

Table \ref{tab:band_gap} summarizes the experimental and theoretical
values of the band gap of MoS$_2$. Early photoluminescence experiments
\cite{Mak2010,Splendiani2010} had inferred a direct band gap of about
1.9 eV for MoS$_2$. More recently, it has been suggested that this
value is actually the result of excitonic states and hence corresponds
to the optical gap rather than the actual direct gap between the
single-particle VB and CB \cite{Qiu2013}. Scanning Tunneling
Spectroscopy (STS) measurements revealed that the band gap of MoS$_2$
is 2.15 eV \cite{Zhang_exciton2014}. Given the optical gap of about
1.9 eV, the latter value is quite consistent with both theoretical and
experimental values of the exciton binding energy, which fall in the
range 0.28--0.33 eV according to theory \cite{Zhang2014,Wu2015} and
are either 0.44 eV \cite{Hill2015} or 0.22 eV \cite{Zhang_exciton2014}
as deduced from experiments. Traditional DFT functionals based on the
local density approximation (LDA) and on the generalized gradient
approximation (GGA), not surprisingly, underestimate this band gap
\cite{Lebegue2009,Torres2015,Mann2014}, while the more advanced GW approach tends
to overestimate it \cite{Qiu2013}. The HSE06 functional
\cite{Heyd2003,erratumHeyd2003}, on the other hand, provides so far
the best agreement \cite{Kang2013} with the STS result for this gap
\cite{Zhang_exciton2014}.

\begin{figure}[h]
\begin{centering}
  \includegraphics[width=0.6\columnwidth]{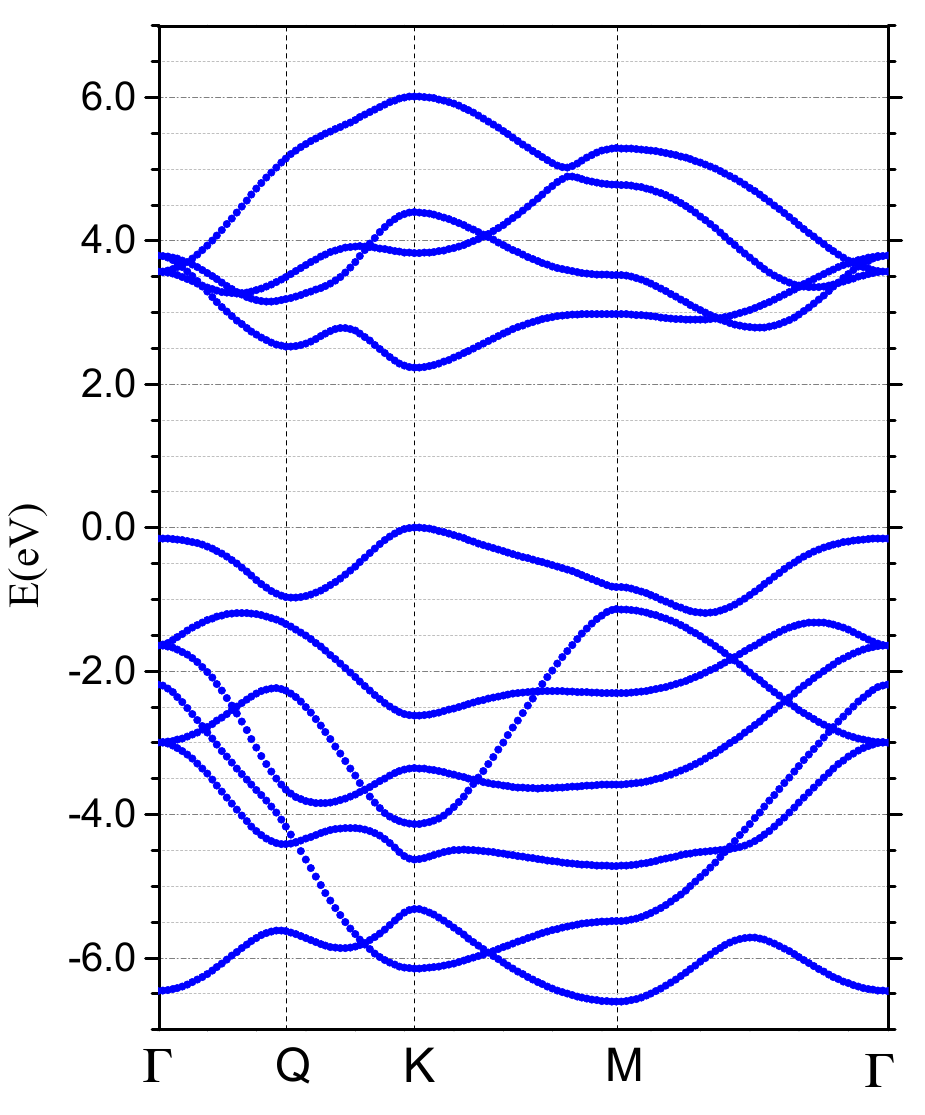} 
  \par\end{centering}
\protect\caption{The DFT-HSE06 band structure of MoS$_2$ near the gap
  region. See text for details.}
\label{fig:DFT-energy-bands}
\end{figure}

In this work, we have therefore chosen the DFT-HSE06 band structure as
reference for our fitting procedures. Our DFT-based electronic band
structure calculations are carried with the HSE06 functional using the
supercell method with a plane-wave basis set (cutoff energy of 500 eV)
and the projector-augmented wave (PAW) technique \cite{PAW1994,
  Kresse1999}, as implemented in the Vienna ab-initio Simulation
Package (VASP) \cite{Kresse1993,kresse1996}. We use a supercell
consisting of a MoS$_2$ layer with an experimental lattice parameter
value of 3.16 \AA~at its center and a vacuum of 15 \AA~ to minimize
the interaction between normal periodical images. The structure is
optimized using the GGA approximation with the Perdew-Burke-Ernzerhof
(PBE) parameterization \cite{Perdew1996}. The Brillouin zone is
sampled by a $18 \times 18 \times 1$ mesh. In calculations including
spin-orbit-coupling, we sample the Brillouin Zone with a $9 \times 9
\times 1$ $k$-point mesh to reduce computational cost. The electronic
band structure along the $\Gamma$--$K$--$M$--$\Gamma$ directions is
calculated with 149 $k$-points and then projected onto every orbital
of each atom to resolve the symmetry characters of the corresponding
wave-functions. The resulting band structure for a MoS$_{2}$ monolayer
is shown in Fig. \ref{fig:DFT-energy-bands}.

%End modification from Duy

%%%%% FIGURE 3
%--------------------------------------------------------------------------
%--------------------------------------------------------------------------

Let us summarize the main features near the high-symmetry points of
the Brillouin zone \cite{Kormanyos2013,Zahid2013}:

\begin{itemize}
\item {\it $K$ point} -- The calculated (DFT-HSE06) band gap
  $E_G=2.23$ eV is located at the $K$ point. This result is good
  agreement with experimental value of 2.15 eV
  \cite{Zhang_exciton2014}. Electron-hole symmetry is clearly absent:
  While the effective mass $m_{e}$ in the CB is nearly isotropic, in
  the VB band it is characterised by a strong trigonal warping. Spin
  splitting is present in both CB and VB, but the splitting of the VB
  states is much larger. The VB spin splitting at the $K$ point is
  experimentally found to be around $145$~meV. Here we only consider
  spin-orbit coupling up to first order in the coupling and hence
  disregard the spin splitting in the CB at the $K$ point. Higher
  order SOC effects have been studied in
  Refs.~\cite{Kosminder2013,Ferreiros2014}.

\item {\it $Q$ point} -- This point, signaling a local minimum in the
  CB band along the straight line connecting $\Gamma$ and $K$ points,
  has recently received increased attention due to its relevance for
  transport properties, since the energy minimum $E_Q$ is close to the
  bottom of the CB \cite{Hawrylak2012,Kormanyos2013}. From our
  DFT-HSE06 calculations, before including SOC, we estimate this
  energy difference to be $\Delta E \approx 0.3$ eV. It is noteworthy
  that the CB at the $Q$ point moves down in energy in multilayer
  systems. As discussed in Ref.~\cite{Cap2014}, phonon-limited
  mobility depends quite sensitively on this energy separation. At the
  $Q$ point, the CB is characterized by a spin splitting of $91$~meV
  and the effective mass has an ellipsoidal shape
  \cite{Hawrylak2012}. The $Q$ point is located close to the mid point
  between $\Gamma$ and $K$ points.

\item {\it $\Gamma$ point} -- This point lies close to the top of the
  valence band. According to our DFT-HSE06 calculations, before
  including SOC, its energy difference to the $K$ point is very small,
  namely, $\Delta E \approx 0.15$ eV. Hence, in hole-doped samples
  states both the $K$ and $\Gamma$ points will contribute to the
  electronic transport.
\end{itemize}

The orbital composition is of fundamental importance for building of
any tight-binding model. As already found in literature
\cite{Cap2013}, $d_{xy}, d_{x^{2}-y^{2}},d_{z^{2}}$, and $p_{x,y}$ are
the most important orbitals to describe the valence and the conduction
bands. It is worth to stress that $d_{yz}$ and $d_{xz}$ have a
dominant contribution at the $\Gamma$ point, and $p_{z}$ gives an
important contribution to both $\Gamma$ and $Q$ points. Thus, for a
comprehensive description of the CB and VB along the Brillouin zone,
one needs to consider all these orbitals. Tables
\ref{tab:orb_comp_G_DFT} and \ref{tab:orb_comp_K_DFT} show the
relative contribution from each orbital at the points $\Gamma$ and
$K$, as provided by our DFT-HSE06 calculations. These results serve
not only to justify the choice of the relevant orbitals of the
atomistic model, but also help in finding the right constraints to
optimize the tight-binding parameters.

%---------- TABLE 1 [IOP format] -----------------------------------------
\begin{table}[h]
\centering
\caption{Density functional theory (DFT-HSE06) orbital composition at
  the $\Gamma$ point for different bands. Absence of an entry
  indicates zero contribution.}
\label{tab:orb_comp_G_DFT} 
\begin{tabular}{cccccccccc}
\br
band     & band    &&&&&&&&\\
number & energy (eV) &  $p_{y}$ & $p_{z}$ & $p_{x}$ & $d_{xy}$ & $d_{yz}$ & 
$d_{z^{2}}$ & $d_{xz}$ & $d_{x^{2}}$ \\
\mr
6  & -7.571   & & 0.252 &  &  &  & 0.199 &  &  \\
7  & -4.105   & 0.115 &  & 0.127 & 0.309 &  &  &  & 0.279 \\
8  & -4.105  & 0.127 &  & 0.115 & 0.279 &  &  &  & 0.309 \\
9 & -3.303   &  & 0.558 &  &  &  &  &  &  \\ 
10 & -2.753   & 0.009 &  & 0.313 &  & 0.015 &  & 0.502 &  \\
11 & -2.753   & 0.313 &  & 0.009 &  & 0.502 &  & 0.015 &  \\
12 VB & -1.262   &  & 0.141 &  &  &  & 0.596 &  &  \\
13 CB & 2.457    & 0.302 &  & 0.045 &  & 0.327 &  & 0.049 &  \\
14 & 2.457    & 0.045 &  & 0.303 &  & 0.048 &  & 0.327 &  \\
15 & 2.678    & 0.083 &  & 0.390 & 0.249 &  &  &  & 0.053 \\
16 & 2.678   & 0.390 &  & 0.083 & 0.053 &  &  &  & 0.249 \\
\br
\end{tabular}
\end{table}
%----------- end TABLE 1 [IOP format] ----------------------------------

%-------- TABLE 2 [IOP format] -----------------------------------------
\begin{table}[h]
\centering
\caption{Density functional theory (DFT-HSE06) orbital composition at
  the $K$ point for different bands. Absence of an entry indicates
  zero contribution.}
\label{tab:orb_comp_K_DFT}
\begin{tabular}{ccccccccccc}
\br
band     & band    &&&&&&&&&\\
number & energy (eV) &  $p_{y}$ & $p_{z}$ & $p_{x}$ & $d_{xy}$ & $d_{yz}$ & 
$d_{z^{2}}$ & $d_{xz}$ & $d_{x^{2}}$ \\
\mr
6  & -7.259 &   0.155 &  & 0.155 &  & 0.184 &  & 0.184 &  \\
7  & -6.427  & 0.178 &  & 0.178 &  &  & 0.135 &  &  \\
8  & -5.742  & 0.231 &  & 0.231 & 0.034 &  &  &  & 0.034 \\ 
9 & -5.244  & 0.034 & 0.359 & 0.034 & 0.105 &  &  &  & 0.105 \\ 
10 & -4.466 &   0.230 & 0.129 & 0.230 &  &  &  &  &  \\ 
11 & -3.734   & 0.416 &  &  & 0.140 &  & 0.140 &  \\ 
12 VB & -1.111   & 0.065 &  & 0.065 & 0.345 &  &  &  & 0.345\\ 
13 CB & 1.120  & 0.034 &  & 0.034 &  &  & 0.753 &  & \\ 
14 & 2.718 &   0.068 &  & 0.068 &  & 0.248 &  & 0.248 & \\ 
15 & 3.284  & 0.016 & 0.153 & 0.016 & 0.327 &  &  &  & 0.327\\ 
16 & 4.899  &  & 0.188 &  &  & 0.303 &  & 0.303 &  \\
\br
\end{tabular}
\end{table}
%------------ end TABLE 2 [IOP format] ----------------------------------

%%%%%%%%%%%%%%%%%%%%%%%%%%%%%%%%%%%%
%\input {3model}
%  Model
%
\section{Model}
\label{sec:3model}
%%%%%%%%%%%%%%%%%%%%%%%%%%%%%%%%%%%%

Let ${\bf r}_i$ denote the Mo atom location in the $i$th unit
cell. Following Cappelluti and collaborators \cite{Cap2013,Cap2014},
we consider a tight-binding model with five ${d}$ orbitals in the Mo
atom, namely,
$$
%%\begin{eqnarray}  
|{\bf r}_i;d_{0}\rangle = |d_{3z^{2}-r^{2}}\rangle, \ \
|{\bf r}_i;d_{1}\rangle = |d_{x^{2}-y^{2}}\rangle,  \ \
%%\nonumber\\
|{\bf r}_i;d_{2}\rangle = |d_{xy}\rangle, \ \ % \quad\quad\;\;\,
|{\bf r}_i;d_{3}\rangle = |d_{xz}\rangle, \ \ % \quad \quad
|{\bf r}_i;d_{4}\rangle = |d_{yz}\rangle,
%%\end{eqnarray} 
$$
and six $p$ orbitals for the S atoms, three for the top $t$ ($+$) and
three for the bottom $b$ ($-$) layers,
$$
%%\begin{eqnarray} 
|{\bf r}_i+{\bm\delta}_{1\pm};p_{1}\rangle = |p_{x}^{t,b}\rangle, \ \
|{\bf r}_i+{\bm\delta}_{1\pm};p_{2}\rangle = |p_{y}^{t,b}\rangle, \ \
|{\bf r}_i+{\bm\delta}_{1\pm};p_{3}\rangle = |p_{z}^{t,b}\rangle.
%%\end{eqnarray} 
$$
Starting with this basis we can define on-site energies and hopping
amplitudes and write down a tight-binding Hamiltonian. Hereafter we
assume that this basis is orthogonal.

The tight-binding Hamiltonian $\mathcal{H}$ contains Mo--S and S--S
nearest-neighbor hopping terms (in the same unit cell), as well as
Mo--Mo and S--S next-to-nearest-neighbor ones (in adjacent cells).
Each Mo has six S nearest neighbors. while the next-to-nearest
neighbor hoppings connect 6 atoms of the same kind, see
Fig.~\ref{fig:Top-view-of reticolo}. Overall, there is a total of 25
hopping matrix elements inside the unit cell and between the unit cell
and the adjacent cells.

The hopping amplitudes are written in terms of Slater-Koster (SK)
parameters \cite{Slater-Koster}. We incorporate the $x$ and $z$
reflection symmetries in the construction of the basis, when
applicable, to reduce the number of terms. We refer to
\ref{subsec:Appendix A} for a detailed description of the
tight-binding Hamiltonian and the transfer integrals. There, we also
provide expressions for the hopping amplitudes in terms of the SK
integrals $V_{pd\sigma}, V_{pd\pi}, V_{dd\sigma}, V_{dd\delta},
V_{dd\pi}, V_{pp\sigma}$, and $V_{pp\pi}$. This allows for a
significant reduction in the number of fitting parameters of the
model.

To find the energy bands we solve the eigenvector equation that, in
the Bloch momentum representation, reads
\begin{equation}
  \mathcal{H}|{\bf k}\rangle = E_{\sigma}({\bf k})|{\bf k}\rangle,
\label{eq:schrodinger}
\end{equation}
where the eigenstates $|{\bf k}\rangle$ are expressed in terms of the
three-atom basis, namely,
\begin{eqnarray}
|{\bf k}\rangle  =  \sum_{{\bf r}_i} e^{i{\bf k}\cdot{\bf r}_i}&&
\left[ \sum_{\mu=0}^{4} \alpha_{{\bf k},\mu} |{\bf r}_i; d_{\mu}
  \rangle + \right. \nonumber \\  & & \left. \;\;\,\sum_{\mu=1}^{3} \left(
  \beta_{{\bf k},\mu}|{\bf r}_i + {\bm \delta}_{1-}; p_{\mu}, \rangle
  + \tau_{{\bf k},\mu\sigma}| {\bf r}_i + {\bm \delta}_{1+};p_{\mu}
  \rangle \right) \right].
\end{eqnarray} 

For the purpose of implementing the eigenvalue equation we project the
vector $\mathcal{H}|{\bf k}\rangle$ onto the three-atom basis and
write
\begin{equation}
\label{eq:H-unsymmetrized}
\left(\begin{array}{ccc}
h^{\rm{Mo}}+V & T^{t} & T^{b}\\
(T^{t})^{\dagger} & h^{{\rm S}} + U & S\\
(T^{b})^{\dagger} & S & h^{{\rm S}}+U
\end{array}\right)
\left(\begin{array}{c}\alpha\\ \tau\\ \beta \end{array}\right) =
E\left(\begin{array}{c}\alpha\\ \tau\\ \beta \end{array}\right),
\end{equation}
where have omitted, for the moment, the spin indices. Explicit
expressions for the block matrices $h^{\rm Mo}, h^{\rm S}, T^t, T^b, S,
U,$ and $V$ are given in \ref{subsec:Appendix A}. The matrices $S, U$
and $V$ are real and symmetric.

The secular equation (\ref{eq:H-unsymmetrized}) is sufficient for a
numerical evaluation of the band structure. Nonetheless, it is
convenient to use symmetry arguments to reduce the the size of the
matrices to be diagonalized, allowing us to express analytically the
gap and other features of the band structure at $k$ points of
interests.

Let us introduce the symmetric and anti-symmetric components
\begin{equation}
\theta_{\bf{k},\nu} = \frac{1}{\sqrt{2}}
(\tau_{\bf{k},\nu}+\beta_{\bf{k},\nu})
\end{equation}
and
%\quad \mbox{and} \quad
%
\begin{equation}
\phi_{\bf{k},\nu} = \frac{1}{\sqrt{2}}
(\tau_{\bf{k},\nu}-\beta_{\bf{k},\nu}),
\end{equation}
that allow us to write the Hamiltonian in the matrix form
\begin{equation}
\left(\begin{array}{ccc}
h^{\rm{Mo}} & T^{E} & T^{O}\\
T^{E\dagger} & h_{}^{{Sp}} & 0\\
T^{O\dagger} & 0 & h^{{S}m}
\end{array}\right)\left(\begin{array}{c}
\alpha\\
\theta\\
\phi
\end{array}\right)=E\left(\begin{array}{c}
\alpha\\
\theta\\
\phi
\end{array}\right).
\end{equation}
The eigenvalue problem can be further simplified by rearranging rows
and columns through the transformation $\psi \rightarrow \bar\psi$,
where
\begin{equation}
\psi^T = \left( \alpha_{0}, \alpha_{1}, \alpha_2, \alpha_{3},
\alpha_4, \theta_{1}, \theta_{2}, \theta_{3}, \phi_{1}, \phi_{2},
\phi_{3} \right)
\end{equation}
and
\begin{equation}
\bar{\psi}^T = \left( \alpha_{0}, \alpha_{1}, \alpha_2, \theta_{1},
\theta_{2}, \phi_{3}, \alpha_{3}, \alpha_{4}, \phi_{1}, \phi_{2},
\theta_{3} \right).
\end{equation}
Notice that the first six orbital basis functions are even ($E$) with
respect to a $z$-axis inversion, while the last five are odd ($O$).
Then, the problem is reduced to two decoupled eigenvalue/eigenvector
problems, namely
\begin{equation}
\label{eq:HE-HO}
\left(\begin{array}{cc}
H^{E} & 0\\
0 & H^{O}
\end{array}\right) \psi = E \psi.
\end{equation}
We refer to \ref{subsec:Appendix B} for explicit expressions of the
matrix elements of $H^{E}$ and $H^{O}$.

%%%%%%%%%%%%%%%%%%%%%%%%%%%%%%%%%%%%%%%
% Optimization
%\input {4optimization}
% Optimization
%
\section{Optimization of model parameters}
\label{sec:4optimization}
%%%%%%%%%%%%%%%%%%%%%%%%%%%%%%%%%%%%%%%

Our tight-binding model Hamiltonian has $N_p=12$ fitting parameters,
namely, five on-site orbital energies ($D_0, D_1, D_2, D_p$, and
$D_z$) and seven SK parameters related to hopping ($ V_{pd\pi},
V_{pd\sigma}, V_{pp\sigma}, V_{pp\pi}, V_{dd\sigma}, V_{dd\pi},$ and
$V_{dd\delta}$). These parameters are optimized to reproduce the main
characteristics of the low-energy bands we obtained from DFT-HSE06
calculations.

\begin{figure}[h]
\centering
\includegraphics[width=0.5\columnwidth]{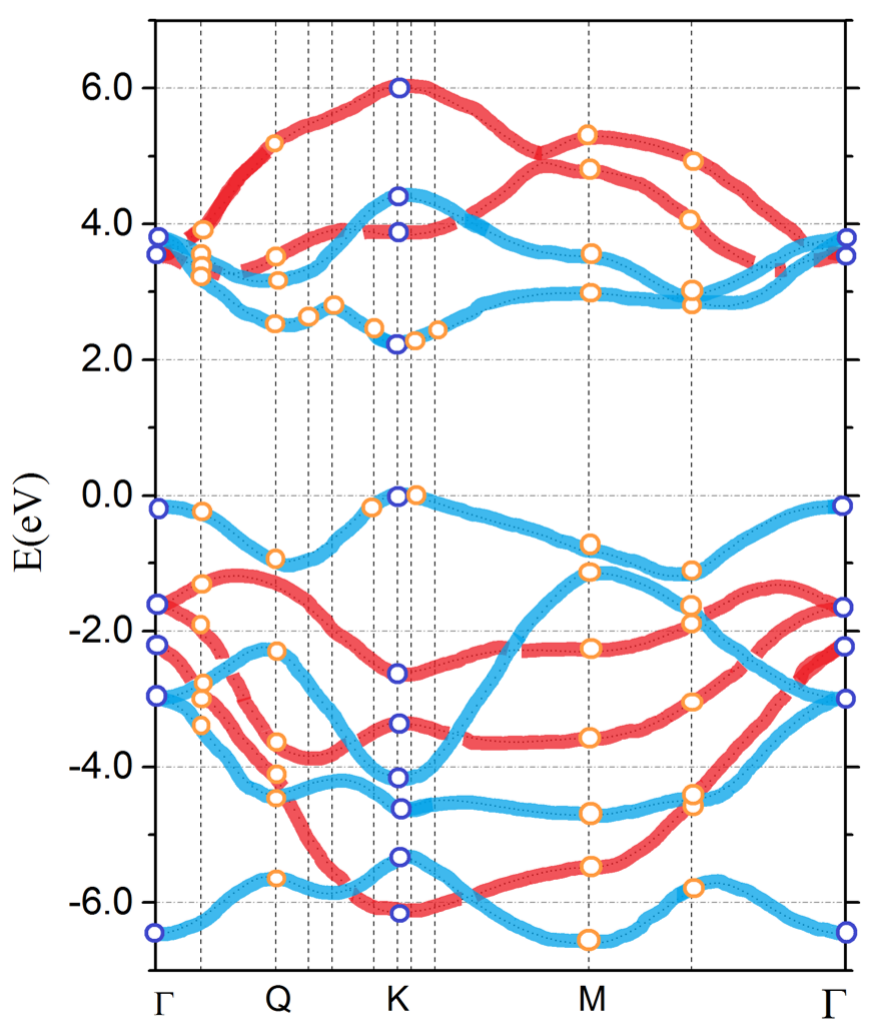}
\caption{Reference DFT-HSE06 band structure with the constraint points
  indicated. Blue circles: analytical constraints. Orange circles:
  numerical constraints. Predominantly even (odd) bands with respect
  to $z$ inversion are shown in blue (red).}
\label{fig:DFT-reference-and}
\end{figure}

Our main goal is to reproduce the energies, orbital composition, and
effective masses of the conduction and valance bands at the $K, Q,$
and $\Gamma$ points. For that purpose we choose a number of
representative $k$-points, shown in Fig.~\ref{fig:DFT-reference-and},
and collect the corresponding band energies $E_n({\bf k})$, where $n$
is the band index, to built the data set to be fitted. To better
describe the conduction and valence energy bands, we give a larger
weight to points $({\bf k}, E_n({\bf k}))$ near the main band gap. In
addition, we take a larger concentration of points around $K, Q,$ and
$\Gamma$ to reproduce the electron effective mass around these high
symmetry-points.

%-----------------------------------------------------------------------------
%-----------------------------------------------------------------------------

We find the optimal tight-binding parameters using the method of least
squares. The data set is built from the band energies $E_{j}^{\rm
  DFT}$, where $j$ labels both the $k$-point and band index (see
Fig.~\ref{fig:DFT-reference-and}), with $j=1,\cdots,N_f$. The
corresponding $\chi^2$-squared function is just a sum of weighted
squared residuals, namely,
\begin{equation}
S({\bf P}) = \sum_{j=1}^{N_f} \frac{ \left[ E_{j}^{\rm tb}({\bf P}) -
    E_{j}^{\rm DFT} \right]^{2}} {\sigma_{j}^2}.
\end{equation}
where $\sigma_j$ is the weight given to the $j$th data set fitting
point, $E_{j}^{\rm tb}({\bf P})$ is the tight-binding energy
corresponding to the $j$th data set fitting point. The vector array
${\bf P}$ of dimension $N_p$ contains the tight-binding parameters to
be minimized. We minimize $S$ with respect to ${\bf P}$ using the
Powell method \cite{Fortran}, that is an efficient method to find the
minimum of a function of several variables without requiring the
computation of its derivatives.

Let us briefly describe the route we follow to approximate the
low-energy band structure by progressively adding data points.

\begin{enumerate}
\item We compare the results obtained at the $K$ and $\Gamma$ points
  using the analytical expressions derived in \ref{subsec:Appendix C}
  to the DFT-HSE06 energy values and their orbital compositions.

\item We consider $k$ points in the vicinity of $K$ and $\Gamma$. The
  weights $\sigma$ are adjusted to decrease the importance of these
  $k$ points as the further away they are from the gap region.
 
\item We consider additional data points to correctly reproduce other
  features of the CB and VB. In particular, we add $k$ points at and
  around the $M$ and $Q$ points, to obtain a correct band energy
  behavior around the main gap over the entire Brillouin zone.
\end{enumerate}

The tight-binding model Hamiltonian decouples into ``even'' bands
(associated to $H_{E}$) and ``odd'' bands (associated to $H_{O}$). The
identification of the parity of the bands at the $\Gamma$ and $K$
points allows us to follow all the bands over the entire Brillouin
zone. Notice that around the main energy gap, bands are mostly even
(except for the CB at the $\Gamma$ point). Let us now explain how we
match DFT-HSE06 band energies with the tight-binding eigenvalues at
the $K$ and $\Gamma$-points.

%_____________________________________________________
\subsection{$K$ point}
\label{sec:Kpoint}

At the $K$ point, the matrices $H^E$ and $H^O$ in Eq.~(\ref{eq:HE-HO})
can be written in block diagonal form, namely, we can break $H^{E}$
into three $2\times2$ diagonal blocks and $H^{O}$ into two $2\times2$
blocks and one $1\times1$ block. The explicit expressions are
presented in \ref{subsec:Appendix C}.

The inspection of the orbital composition given by the DFT-HSE06
calculations (see Table \ref{tab:orb_comp_K_DFT}) allows us to
establish a unique correspondence between pairs of bands and the
$2\times2$ blocks mentioned above. The correspondence is summarized in
Table \ref{tab:Contribution-from-each-gamma-1}.

The identification of the highest $E_+$ (lowest $E_-$) eigenvalue of a
given block with the highest (lowest) energy value of the band with
corresponding orbital composition and parity severely constrains our
model parameters.  By applying this methodology to all diagonal
blocks, we find analytical expressions for the band energies at the
$K$ point, presented in \ref{subsec:Appendix C}.

%------------ TABLE 3   [IOP format] -----------------------------------------
\begin{table}[h]
\centering
\caption{Identification of the tight-binding 2$\times$2 block
  structures and their orbital contributions at the $K$ point with the
  band numbers and their corresponding DFT-HSE06 energies, given in
  Table \ref{tab:orb_comp_K_DFT}.}
\label{tab:Contribution-from-each-K-1}
\begin{tabular}{cccc}
\br 
block structure  orbitals  & orbital composition & $(E_+,E_-)$ (eV) & band numbers \\
\mr 
$H_E: \alpha_0-   \theta_R$ & $d_{z^{2}}, p_{x}, p_{y}$ &
$(1.120,-6.427)$ & (13CB, 7) \\
$H_E: \alpha_{L2}-\theta_L$ &  $d_{x^{2}-y^{2}}, d_{xy}, p_{x}, p_{y}$ &
$(-1.111,-5,742)$ & (12VB, 8) \\
$H_E: \alpha_{R2}-\phi_3$   & $d_{x^{2}-y^{2}}$, $d_{xy}$, $p_{z}$ &
$(3.284,-5.244)$ & (15,9) \\
$H_O: \alpha_{R1}-\phi_L$   & $d_{yz}$, $d_{xz}$, $p_{x}$,$p_{y}$ &
$(2.457,-2.753)$ & (14,6) \\
$H_O: \alpha_{L1}-\theta_3$ & $d_{yz}$,  $d_{xz}$, $p_{z}$ &
 $(4.899,-3.734)$ & (16,11) \\
 $H_O: \phi_{R} $  & $p_{x}, p_{y}$ & 
 $-4.466 $ & 10 \\
 \br
\end{tabular}
\end{table}
%----------------- end TABLE 3 ---------------------------------------------

%_____________________________________________________
\subsection{$\Gamma$ point}

At the $\Gamma$ point we can also express the $H$ matrix of
Eq.~(\ref{eq:HE-HO}) in block block diagonal form, breaking it into
five $2\times2$ blocks and one $1\times1$ block.

Here we follow the procedure described in the previous subsection,
Sec.\ \ref{sec:Kpoint}. The differences are due to the distinct
point-group symmetries of the $K$ and $\Gamma$ points. As a
consequence, the orbital compositions of the tight-binding 2$\times$2
blocks considered in this case are not the same as for the $K$ point.
This issue is discussed in \ref{subsec:Appendix C}, where we also
present the analytical derivation of eigenvalues and eigenstates at
the $\Gamma$-point.

Table \ref{tab:Contribution-from-each-gamma-1} presents the
identification of the tight-binding symmetry split 2$\times$2 blocks
with their corresponding DFT-HSE06 bands. It is also noteworthy that,
as presented in Table \ref{tab:orb_comp_G_DFT}, the {\it ad initio}
calculations show that several band energies coincide at the $\Gamma$
point, namely, 7 and 8, 10 and 11, 13CB and 14, and 15 and 16.

%------------ TABLE 4   [IOP format] -----------------------------------------
\begin{table}[ht]
\centering
\caption{Identification of the tight-binding 2$\times$2 block
  structures and their orbital contributions at the $\Gamma$ point
  with the band numbers and their corresponding DFT-HSE06 energies,
  given in Table \ref{tab:orb_comp_G_DFT}. In the last column, cases
  where the bands $n_i,$ and $n_j$ have the same energy at the
  $\Gamma$-point are denoted by $n_{i}$--$n_{j}$.}
\label{tab:Contribution-from-each-gamma-1}
\begin{tabular}{cccc}
\br 
block structure  orbitals  & orbital composition & $(E_+,E_-)$ (eV) & band numbers \\
\mr 
$H_E: \alpha_0-   \phi_3$ & $d_{z^{2}}$,  $p_{z}^{t}$, $p_{z}^{b}$ &
$(-1.262,-7.571)$ & (12VB, 6) \\
$H_E: \alpha_1-\theta_2$ &  $d_{x^{2}-y^{2}}$, $p_{x}^{t}$, $p_{x}^{b}$ &
$(2.678,-4.105)$ & (15-16,7-8) \\
$H_E: \alpha_2-\theta_1$ & $p_{y}^{t}$, $p_{y}^{b}$ &
$(2.678,-4.105)$ & (15-16,7-8) \\
$H_O: \alpha_{3}-\phi_{1}$ & $d_{xz}$,  $p_{x}^{t}$, $p_{x}^{b}$ &
$(2.457,-2.753)$ & (13CB-14,10-11) \\
$H_O: \alpha_{4}-\phi_{2}$  & $d_{yz}$,  $p_{y}^{t}$, $p_{y}^{b}$ &
 $(2.457,-2.753)$ & (13CB-14,10-11) \\
 $H_O: \theta_{3}$  & $p_{z}^{t}$, $p_{z}^{b}$ & 
 $-3.303 $ & 9 \\
 \br
\end{tabular}
\end{table}
%----------------- end TABLE 4 ---------------------------------------------

%%%%%%%%%%%%%%%%%%%%%%%%%%%%%%%%%%%%%%%%%
\section{Eleven-band model: parameters and results}
\label{sec:5-11band}
%%%%%%%%%%%%%%%%%%%%%%%%%%%%%%%%%%%%%%%%%

In this Section we present the main results of our study, namely, the
tight-binding 11-band parametrization and the corresponding band
structure for MoS$_2$. Table \ref{tab:Tight-binding-parameters}
presents the best fitting parameters we obtained using the the
optimization procedure described in Sec. \ref{sec:4optimization}.

Before discussing the results, it worth mentioning that the even-odd
parity symmetry of our tight-binding model prevents a perfect match
with {\it ab initio} calculations. For instance, DFT-HSE06
calculations indicate that the CB and VB are mainly ``even'', but
around the $\Gamma$ point they gain a significant odd
contribution. Despite this proviso, we show that the tight-binding
model reproduces the {\it ab initio} band structure close Fermi energy
with very good accuracy.

Although our tight-binding model contains many adjustable parameters,
the optimization procedure presented in the
Sec. \ref{sec:4optimization} imposes several implicit constraints. In
practise, we find very difficult to obtain a parameter set that
reproduces with high accuracy the position of the energy bands, their
orbital compositions, {\it and} effective masses at the $K, \Gamma$,
and $Q$ points for both CB and VB. For this reason, in Table
\ref{tab:Tight-binding-parameters} we present two parameter sets: one
that reproduces most features of both VB and CB, but does not yield
accurate masses for the VB, and the other that focuses on the VB.

%----------- TABLE 5 [IOP format] -----------------------------------------
\begin{table}[h]
\centering
%\begin{indented}
\caption{
  Tight-binding model parameters obtained by optimization using
  $a=3.16\textrm{\AA}$, $\theta_{B}=0.710$, and $d=2.406$ \AA. The second 
  column gives the best parameter set we obtain to fit both the valence (VB) and
  the conduction (CB) bands, while the third column focuses the optimization on
  the valence band.}
\label{tab:Tight-binding-parameters}
\lineup
\begin{tabular}{ccc}
\br
parameters  & CB-VB optimization (eV) & VB optimization (eV) \\
%& (eV) & (eV) \\
\mr
$D_{0}$ & 0.201 & 0.191\\
$D_{1}$ & -1.563 & -1.599\\ 
$D_{2}$ & -0.352 & 0.081\\ 
$D_{p}$ & -54.839 & -48.934\\
$D_{z}$ & -39.275 & -37.981\\
$V_{pd\pi}$ & 4.196 & 4.115\\
$V_{pd\sigma}$ & -9.880 & -8.963\\
$V_{pp\sigma}$ & 12.734 & 10.707\\
$V_{pp\pi}$ & -2.175 & -4.084\\
$V_{dd\sigma}$ & -1.153 & -1.154\\ 
$V_{dd\pi}$ & 0.612 & 0.964\\  
$V_{dd\delta}$ & 0.086 & 0.117\\
\br
\end{tabular}
\end{table}
%----------------- end TABLE 5 ---------------------------------------------

\begin{figure}[h]
\centering
\includegraphics[width=.5\columnwidth]{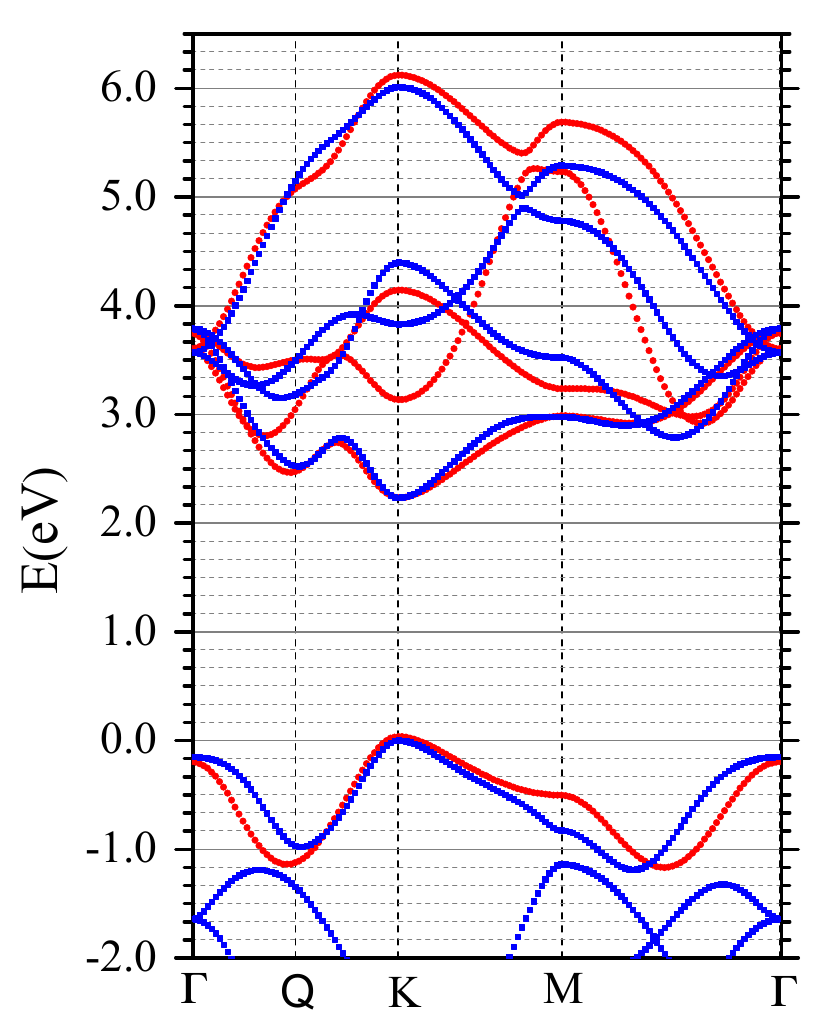}
\caption{Comparison between the band structures obtained with the
  DFT-HSE06 (blue) and with the optimized tight-binding model using
  the parameters from the CB-VB optimization (red) near the gap
  region.}
\label{fig:Fig-a-Comparision_fit} 
\end{figure}

Figure \ref{fig:Fig-a-Comparision_fit} shows the tight-binding band
structure calculated with the VB-CB optimized parameter set given in
Table \ref{tab:Tight-binding-parameters} superposed with the DFT-HSE06
result. We find a very good agreement for the conductance and valence
bands energies. The accuracy of the tight-binding results becomes
increasingly poorer for band energies further away from the gap
region, which is expected given that they were attributed a small
weight in the fitting procedure.

%%%%%%%%%%%%%%%%%%%%%%%%%%%%%%%%%%%
% START FIGURE : dft vs tight-binding
%%%%%%%%%%%%%%%%%%%%%%%%%%%%%%%%%%%

For completeness, in Table \ref{tab:Comparison-orbital} we also
include a comparison of the main orbital composition obtained from
DFT-HSE06 and the tight-binding result. The orbital compositions at
the high symmetry points $\Gamma, K,$ and $Q$ are not equal to those
obtained with the DFT-HSE06, but they show the correct leading $d$ and
$p$ orbitals for both CB and VB. We point out that this is not the
case in the tight-binding parametrization of Ref. \cite{Cap2013},
where several bands near the main energy gap have incorrect
compositions. In particular, at the $K$ point, the correct composition
of the VB appears at a high-energy band, far from the gap. In our
parametrization, out of the 18 points used in the optimization where
analytical expressions where employed, only four yield incorrect
compositions and they are located away from the main gap, at low
energies.

%%%%%%%%%%%%%%%%%%%%%%%%%%%%%%%%%%%%%%%%%
%TABLE: orbital composition check
\begin{table}[h]
\centering
\lineup
\caption{Top: Contribution from each orbital at the $\Gamma$ point using 
  the CB-VB optimization. Bottom: same but at the $K$ point. Omitted 
  orbitals have negligible or null contribution. t-b stands for 
tight-binding model.}
\label{tab:Comparison-orbital}
\begin{tabular}{ccccccc}
\br
band number      &        &        &         &         &           & \\
($\Gamma$ point) & $p_{y}$ & $p_{z}$ & $p_{x}$ & $d_{yz}$ & $d_{z^{2}}$ & $d_{xz}$ \\
\mr
DFT-HSE06 12 VB  &        & 0.141   &        &         & 0.596     &       \\
DFT-HSE06 13 CB  & 0.302  &         & 0.045  &  0.327  &           & 0.049 \\
t-b 12 VB        &        & $1.4\cdot10^{-2}$ &  &   & 0.985 &  \\
t-b 13 CB & $0.11$ &  & $6.6\cdot10^{-6}$ &  0.889 &  & $5.4\cdot10^{-5}$\\
\br
\end{tabular}
\lineup
\begin{tabular}{ccccccc}
\br
band number      &        &        &         &         &          & \\
($K$ point) & $p_{y}$  & $p_{x}$ & $d_{xy}$ &  $d_{z^{2}}$ &  $d_{x^{2}}$ \\
\mr
DFT-HSE06 12 VB & 0.065 &   0.065 & 0.345   &  &   0.345 \\
DFT-HSE06 13 CB & 0.034 &   0.034 &  &   0.753   &  \\
t-b 12 VB & $2.7\cdot10^{-4}$ &   $2.7\cdot10^{-4}$ & 0.499   &  &   0.499 \\ 
t-b 13 CB & $8.9\cdot10^{-3}$ &   $8.9\cdot10^{-3}$ &  &   0.982   &  \\
\br
\end{tabular}
\end{table}
%------------------------------------------------------------------------------

%------------------------------------------------------------------------------
%TABLE: orbital masses check
\begin{table}[h]
\centering
\lineup
\caption{Effective masses (in units of the free electron mass) at the $\Gamma$ and $K$ 
points resulting from the CB-VB and the VB optimization. }
\label{tab:Comparison-masses}
\begin{tabular}{cccc}
\br
      &    HSE06    &  CB-VB & VB  \\
\mr
$\Gamma$ point $m_{e}$ & 0.76 & 0.35 & \\
$\Gamma$ point $m_{h}$ & -2.47 & -0.62 & -2.47\\
\mr
$K$ point  $m_{e}$ & 0.42 & 0.58 & \\
$K$ point $m_{h}$ & -0.47 & -0.61& -0.62\\
\mr
$Q$ point  $m_{e}$ & 0.59 & 0.59& \\
\br
\end{tabular}
\end{table}
%------------------------------------------------------------------------------

\begin{figure}[h]
\centering
\includegraphics[width=.5\columnwidth]{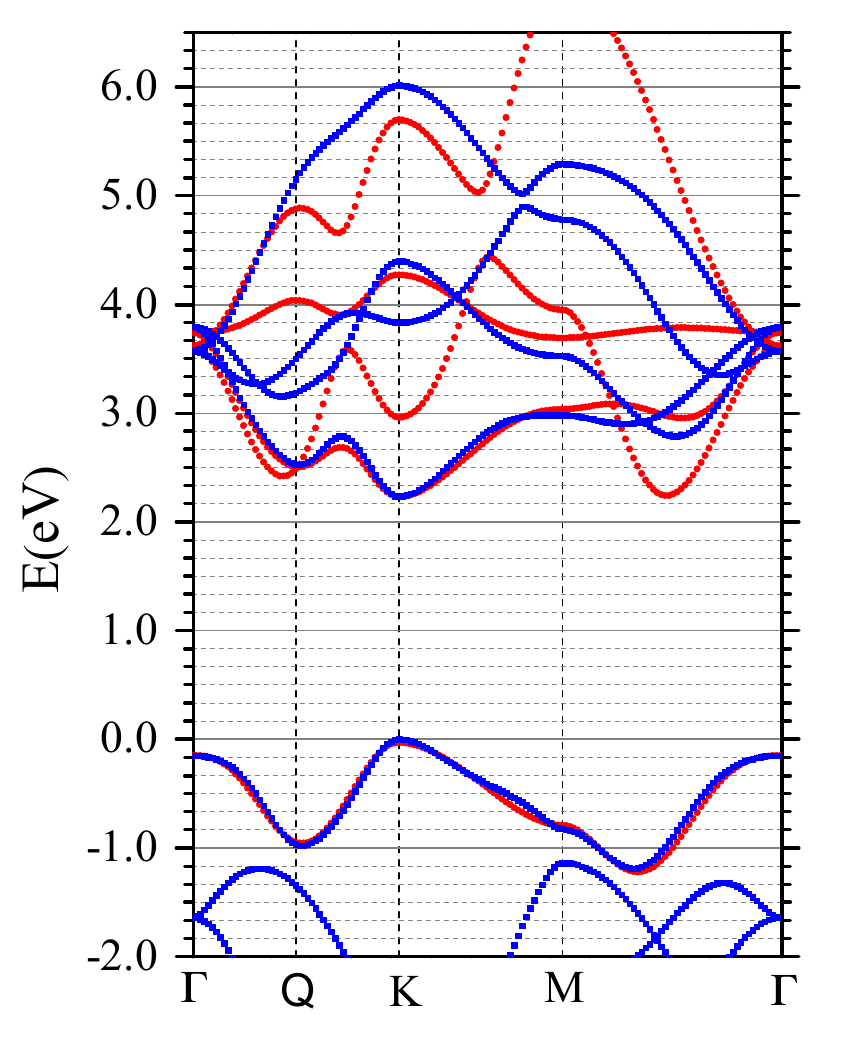}
\caption{Comparison between the band structures obtained with the
  DFT-HSE06 (blue) and with the optimized tight-binding model using
  the parameters from the VB optimization (red) near the gap
  region. The VB optimization focuses on reproducing accurately the
  valence band.}
\label{fig:Fig-a-Comparision_fit-bv} 
\end{figure}

As shown in Table \ref{tab:Comparison-masses}, the effective masses
are also reasonably well described by the CB-VB parametrization for
all three special $k$ points, except for the hole effective mass at
the $\Gamma$ point. We try to circumvent this limitation by performing
another optimization (named VB) with a heavier weight given to the $k$
values near symmetry points at the VB band. The resulting
parametrization describes much more accurately the VB alone, imposing
only few distortions on the CB, as
Fig. \ref{fig:Fig-a-Comparision_fit-bv} reveals. This procedure yields
the VB optimization parameters given in Table
\ref{tab:Tight-binding-parameters} and the orbital compositions and
effective masses presented in Tables \ref{tab:Comparison-orbital-bv}
and \ref{tab:Comparison-masses}, respectively. Most parameter
values are close to those of the global optimization (Table
\ref{tab:Tight-binding-parameters}), while a few differ by more than
25\%. The orbital compositions are nearly identical to those obtained
in the global optimization. The most striking change is in the hole
band mass the $\Gamma$ point, which become essentically identical to
the DFT-HSE06 value.

%%%%%%%%%%%%%%%%%%%%%%%%%%%%%%%%%%%
% START FIGURE : dft vs tight-binding
%%%%%%%%%%%%%%%%%%%%%%%%%%%%%%%%%%

%-------------------------------------------------------------------------------------------------------
%TABLE: orbital composition check
\begin{table}[h]
\centering
\lineup
\caption{Top: Contribution from each orbital at the $\Gamma$ point 
  using the parameters from the VB optimization. Bottom: Same but 
  at the $K$ point. Omitted orbitals have negligible or null 
  contribution. t-b stands for tight-binding model.}
\label{tab:Comparison-orbital-bv}
\begin{tabular}{ccccccc}
\br
band number      &        &        &         &         &           & \\
($\Gamma$ point) & $p_{y}$ & $p_{z}$ & $p_{x}$ & $d_{yz}$ & $d_{z^{2}}$ & $d_{xz}$ \\
\mr
DFT-HSE06 12 VB  &        & 0.141   &        &         & 0.596     &       \\
%DFT-HSE06 13 CB  & 0.302  &         & 0.045  &  0.327  &           & 0.049 \\
t-b 12 VB        &        & $1.2\cdot10^{-2}$ &  &   & 0.988 &  \\
%t-b 13 CB & $11.08\cdot10^{-2}$ &  & $6.6\cdot10^{-6}$ &  0.889 &  & $5.4\cdot10^{-5}$\\
\br
\end{tabular}
\lineup
\begin{tabular}{ccccccc}
\br
band number      &        &        &         &         &          & \\
($K$ point) & $p_{y}$  & $p_{x}$ & $d_{xy}$ &  $d_{z^{2}}$ &  $d_{x^{2}}$ \\
\mr
DFT-HSE06 12 VB & 0.065 &   0.065 & 0.345   &  &   0.345 \\
%DFT-HSE06 13 CB & 0.034 &   0.034 &  &   0.753   &  \\
t-b 12 VB & $6.4\cdot10^{-4}$ &   $6.4\cdot10^{-4}$ & 0.499   &  &   0.499 \\ 
%t-b 13 CB & $8.9\cdot10^{-3}$ &   $8.9\cdot10^{-3}$ &  &   0.982   &  \\
\br
\end{tabular}
\end{table}
%--------------------------------------------------------------------------------------------------

%%%%%%%%%%%%%%%%%%%%%%%%%%%%%%%%%%%%%%%%%%%%%%
%% Optimization easier -just tv
%\input {6easier}
\section{Simplified model}
\label{sec:6easier}
%%%%%%%%%%%%%%%%%%%%%%%%%%%%%%%%%%%%%%%%%%%%%%

The tight-binding model we have developed provides an accurate
description of the main features of the CB and VB at the expense of
involving a relatively large number of orbitals and fitting
parameters. It has already been shown by Liu and coworkers
\cite{Liu2013} that using just three orbitals for the Mo atom and
including only the hopping amplitudes between Mo atoms in plane up to
first neighbours is sufficient to open a band gap. With this in mind,
we explored whether it is possible to neglect some hopping amplitudes
in our tight-binding model and still obtain a reasonable description
of the electronic structure near the band gap region. Keeping only the
hopping amplitudes between Mo atoms turns out to be insufficient, as
it preserves a large amount of degeneracy in the bands. Adding the
hopping amplitudes between Mo and neighboring S atoms, without
including the hopping amplitudes between S atoms, yields reasonable
results. On the other hand, keeping exclusively the Mo--S hoppings
does not yield a band gap. In matrix format, this simplified
tight-binding model yields the eigenvalue/eigenvector problem
\begin{equation}
\left(\begin{array}{ccc}
h^{\rm{Mo}}+V & T^{t} & T^{b}\\
(T^{t})^{\dagger} & h^{{S}} & 0\\
(T^{b})^{\dagger} & 0 & h^{{S}}
\end{array}\right)\left(\begin{array}{c}
\alpha\\
\tau\\
\beta
\end{array}\right)=E\left(\begin{array}{c}
\alpha\\
\tau\\
\beta
\end{array}\right).
\end{equation}

\begin{figure}[h]
\centering
\includegraphics[width=0.5\columnwidth]{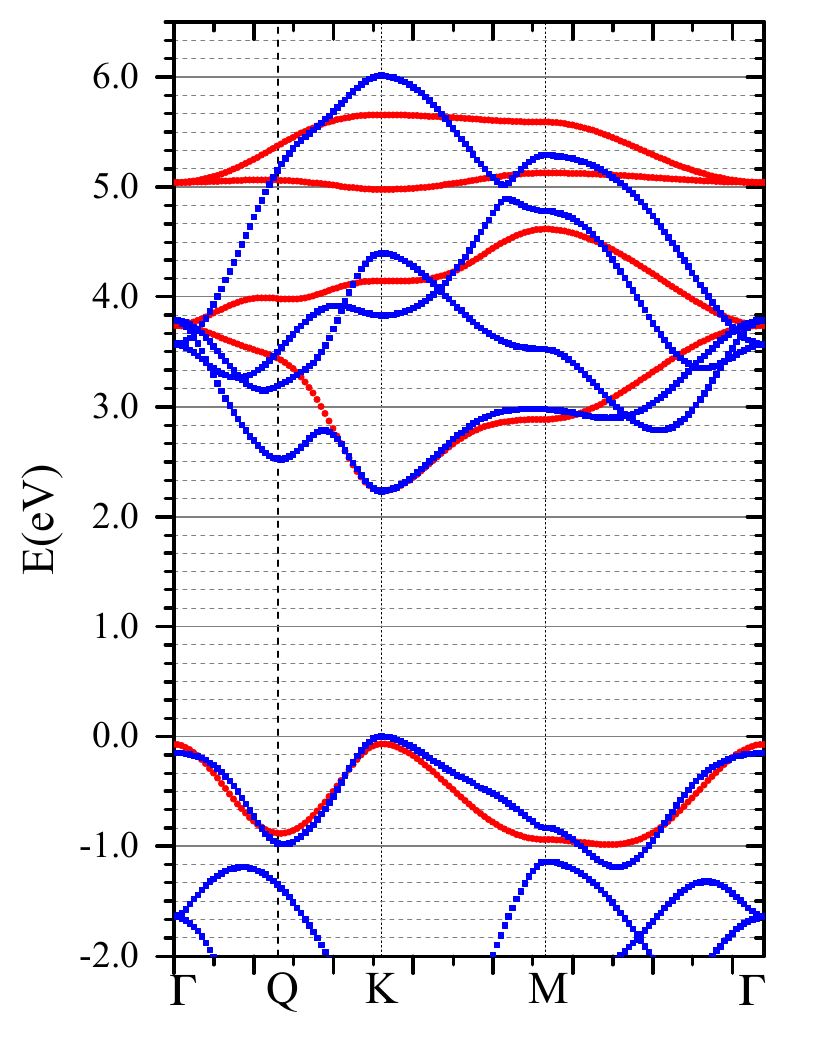}
\caption{Comparison between the DFT-HSE06 band structure (blue points)
  and the best fit to the simplified tight-binding model (red
  points).}
\label{fig:Fig-a)-Comparision_fit_easy} 
\end{figure}

The number of fitting parameters is reduced to from 12 to
10. Symmetries can be fully exploited to break the diagonalization
problem into smaller ones, as done previously. After optimization
against the DFT-HSE06 band structure, we obtain the values for the
fitting parameters listed in Table
\ref{tab:Tight-banding-parameters-easy}. The resulting band structure
is shown in Fig. \ref{fig:Fig-a)-Comparision_fit_easy} superposed with
the DFT-HSE06 band structure. We note that we were able to reproduce
quite well the the entire VB, while the CB is well reproduced just
around the $K$ point, missing the correct behaviour aroung the $Q$ and
$\Gamma$ points. Therefore, the simplified model is somewhat limited
in its applicability. It is suitable for the hole-doped region when
the Fermi energy is brought to the top of the VB. It also provides a
good description of the system when there is weak electron doping.

%--------------- TABLE 6 [IOP format] -----------------------------------------
\begin{table}[h]
\centering
\caption{Parameters for the simplified tight-binding model.}
\label{tab:Tight-banding-parameters-easy}
\begin{tabular}{cc}
\br
parameter & value (eV)\\
\mr
$D_{0}$ & -11.683\\
$D_{1}$ & -208.435\\ 
$D_{2}$ & -75.942\\ 
$D_{p}$ & -23.761\\ 
$D_{z}$ & -35.968\\
$V_{pdp}$ & 1.318\\
$V_{pds}$ & -56.738\\
$V_{dds}$ & -2.652\\ 
$V_{ddp}$ & 1.750\\ 
$V_{ddd}$ & 1.482\\
\br
\end{tabular}
\end{table}
%--------------------- end TABLE 6 ---------------------------------------------

%%%%%%%%%%%%%%%%%%%%%%%%%%%%%%
% START FIGURE : dft vs tight-binding
%%%%%%%%%%%%%%%%%%%%%%%%%%%%%%

%%%%%%%%%%%%%%%%%%%%%%%%%%%%%%%%%%%%%%%%
%
% SECTION: Spin-orbit interaction
%
%\input {7spin}
\section{The effect of spin-orbit interaction}
\label{sec:7spin}
%%%%%%%%%%%%%%%%%%%%%%%%%%%%%%%%%%%%%%%%

Due to the broken lattice inversion symmetry, strong spin-orbit
interactions split the spin-degenerate valence bands in MoS$_{2}$
monolayers as well as in other group VI dichalcogenides. The
spin-orbit coupling in this case is due to the Dresselhaus
mechanism. Interestingly, the spin splitting in inequivalent valleys
must be opposite, as imposed by time-reversal symmetry. As mentioned
in Sec. \ref{sec:1introduction}, these features open interesting
possibilities for the control of spins and valleys in these 2D
materials
\cite{Geim2013,WangKis2012,Butler2Dmaterials,Radisavljevic2011}.

Let us focus on the large spin-splitting at the $K$ point of the
VB. Its origin is qualitatively well understood: The valence band
states are mostly made of $d_{xy}$ and $d_{x^{2}-y^{2}}$ orbitals with
$l=2$ and $m=\pm2$. Therefore, the $L_{Mo}^{z}S_{Mo}^{z}$ component of
the SOC naturally gives a valley-dependent splitting of the bands. In
contrast, the dominant contribution of the CB lowest energy state
comes from the $d_{z^{2}}$ orbital with $l=2$ and $m=0$, which cancels
the spin-orbit splitting. These arguments agree with the quantitative
analysis presented in Ref.~\cite{Kosminder2013}

A complete tight-binding model that accounts for the effect of SOC
over the entire Brillouin zone, including explicitly the $p$-orbitals
of the chalcogen atoms, and taking into account the correct orbital
composition of the main bands, is lacking.

In this Section, we present an extension of our tight-binding model
that includes the effect of an atomic spin-orbit coupling on all the
atoms. For that purpose, we follow the formulation presented in
Ref. \cite{Cap2014}. Our starting point is the 11-band tight-binding
spinless model derived earlier, with the Hamiltonian expressed in the
appropriate symmetrized form, namely, where the block Hamiltonians
$H_{E}$ and $H_{O}$ appear explicitly. The spin-orbit coupling term is
inserted in the Hamiltonian by means of a pure intra-atomic spin-orbit
interaction acting on all the atoms, explicitly given by
\begin{equation}
H_{\rm SO} = \sum_{a}\frac{\lambda_{a}}{\hbar}\, {\bf L}_{a} \cdot
{\bf S}_{a},
\end{equation}
where $\lambda_{a}$ is the intrinsic effective SOC constant for an $a$
atom (Mo o S), ${\bf L}_{a}$ is the atomic orbital angular momentum
operator, and ${\bf S}_{a}$ is the electronic spin operator. Hence,
\begin{equation}
\label{eq:Hspinorbit}
H=\left(\begin{array}{cccc}
H_{E}\\
 & H_{O}\\
 &  & H_{E}\\
 &  &  & H_{O}
\end{array}\right)
+\left(\begin{array}{cccc}
M_{EE}^{\uparrow\uparrow} &  &  & M_{EO}^{\uparrow\downarrow}\\
 & M_{OO}^{\uparrow\uparrow} & M_{OE}^{\uparrow\downarrow}\\
 & M_{EO}^{\downarrow\uparrow} & M_{EE}^{\downarrow\downarrow}\\
M_{OE}^{\downarrow\uparrow} &  &  & M_{OO}^{\downarrow\downarrow}
\end{array}\right)
\end{equation}
The matrix elements of $M$ are straightforward to obtain and depend on
the SOC parameters $\lambda_{\rm Mo}$ and $\lambda_{\rm S}$. The
explicit form of the $M$ matrices can be found in
Ref.~\cite{Cap2014}. We note that in Eq. (\ref{eq:Hspinorbit}) both
diagonal and off-diagonal (spin-flip) terms are taken into
account. However, an analysis in Ref. \cite{Cap2014} indicates that
spin-flip terms have a negligible contribution and could be dropped.

\begin{figure}[h]
\centering
\includegraphics[width=0.5\columnwidth]{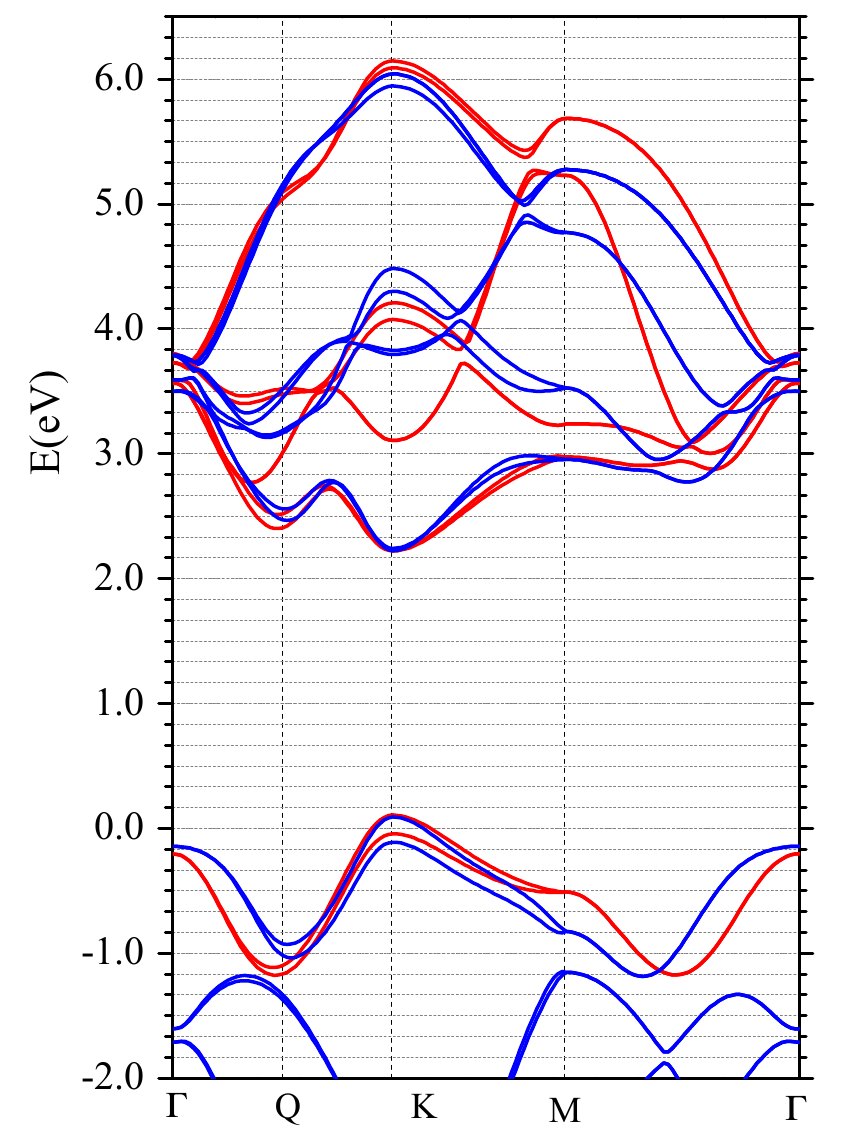}
\caption{Comparison between the DFT-HSE06 spin-resolved band structure
  (blue points) and the best-fit tight-binding model (red points). The
  spin splitting is due to inclusion of spin-orbit coupling.}
\label{fig:Comparision_fit}
\end{figure}

We use DFT-HSE06 to estimate the splittings due to spin-orbit coupling
and obtain $\Delta^{K}=202$ meV at the $K$ point of the VB. This value
is higher than the experimental one \cite{Miwa2015}, $\Delta_{\rm
  exp}^{K}= 145\pm4$ meV. This is a known limitation of the HSE06
functional. The strength of this functional relies on its accuracy to
predict the band gap of numerous materials, including MoS$_2$, where
traditional DFT calculations (LDA or GGA) give significantly
understimated results. Using the SOC values $\lambda_{\rm Mo}=86$
meV and $\lambda_{\rm S}=0.52$ meV, which were obtained from a
tight-binding parameter fit to maximally localized Wannier orbitals
and to DFT calculations \cite{Kosminder2013}, we find $\Delta^{K}=173$
meV. A better result is obtained by adopting the SOC parameters of
Cappelluti {\sl et al.} \cite{Cap2014}, namely, $\lambda_{\rm
  Mo}=75$ meV and $\lambda_{\rm S}=0.52$ meV. Inserting these values
into our tight-binding formulation results instead in $\Delta^{K}=151$
meV, which is in good agreement with the experimental value. Thus we
present in Fig.~ \ref{fig:Comparision_fit} our results for the
spin-resolved band structure based on this choice of SOC parameters.

%%%%%%%%%%%%%%%%%%%%%%%%%%%%%%
% START FIGURE : dft vs tight-binding SPLITTED
%
%
%%%%%%%%%%%%%%%%%%%%%%%%%%%%%%

%%%%%%%%%%%%%%%%%%%%%%%%%%%%%%%%%%%%%%
%
% SECTION: Conclusions
%
%\input {8conclusion}
\section{Conclusions}
\label{sec:8conclusion}
%%%%%%%%%%%%%%%%%%%%%%%%%%%%%%%%%%%%%%

In conclusion, in this paper we provide a suitable and straightforward
tight-binding model for a monolayer dichalcogenides, focusing our
attention on MoS$_{2}$. We show that this model reproduces rather well
the structure near the main energy gap provided by an accurate DFT
band structure calculation based on the HSE06 functional. It also
reproduces the correct orbital composition of the bands. A fundamental
ingredient in obtaining this result is the use of an optimization
process that makes use of analytical expressions of the energy bands
at symmetry points. In the constructing of our model we exploited the
decoupling that exists between even and odd bands upon $z$
inversion. Around the main gap, the bands are primarily even. Overall,
the model yields 11 bands in the absence of spin-orbit coupling and
involves 12 fitting parameters. We provide two parametrizations for
this case: one that is suitable for both conduction and valence bands
(but less accurate for the valence band), and another that gives a
very accurate description of important features of the valence band,
such as the effective mass. When spin-orbit coupling is added, the
number of fitting parameters jumps to 14. Our choice of parameters in
this case yields a spin splitting of the valence band in good
agreement with experimental values.

We also investigate the possibility of turning off some hopping
amplitudes in our model to reduce the number of parameters to 10 in
the absence of spin-orbit coupling. The simplified model is suitable
for describing the hole-doped region or when one is only interested in
the region around the $K$ point.

The present work provides a sound starting point for any further
investigation of electronic transport properties of single-layer
semiconductor transition-metal dichalcogenides, or any other
investigation that relies heavily on an accurate energy level
positioning {\it and} wave function composition.

%%%%%%%%%%%%%%%%%%%%%%%%%%%%%%%%%%%%%%
%  Acknowledgements
\ack We would like to thank Nuno Peres and Marcus Moutinho for helpful
discussions. This work was supported by the Brazilian funding agencies
CNPq, CAPES, FAPERJ, and the Ci\^encia sem Fronteiras program.
D.L. and T.S.R. are supported in part by the DOE grant
DE-FG02-07ER46354.
%Duy add the last line to ack
\appendix

%%%%%%%%%%%%%%%%%%%%%%%%%%%%%%%%%%%%%
\section{Tight-binding model and Slater-Koster parameters}
\label{subsec:Appendix A}
%%%%%%%%%%%%%%%%%%%%%%%%%%%%%%%%%%%%%

The tight-binding model is defined by the following second-quantized
Hamiltonian (spin indices have been omitted):
\begin{eqnarray}
\mathcal{H} & = & \sum_{{\bf r}_i} \sum_{\mu}
\varepsilon_{\mu}^{\rm{Mo}}\, d_{i,\mu}^{\dagger}\, d_{i,\mu}
\nonumber \\ & & + \sum_{{\bf r}_i} \sum_{\nu} \varepsilon_{\nu}^{S}
\left[ \left( p_{i,\nu}^{t} \right)^{\dagger} p_{i,\nu}^{t} + \left(
  p_{i,\nu,\sigma}^{b} \right)^{\dagger}\, p_{i,\nu}^{b} \right]
\nonumber \\ & & + \sum_{{\bf r}_i} \sum_{\mu,\nu} \left[
  t_{\mu\nu}^{t}\, d_{i,\mu}^{\dagger}\, p_{i,\nu}^{t} +
  t_{\mu\nu}^{b}\, d_{i,\mu}^{\dagger}\, p_{i,\nu}^{b} + {\rm H.c.}
  \right] \nonumber \\ & & + \sum_{{\bf r}_i} \sum_{{\bf r}_j={\bf
    r}_i+{\bf R}_1-{\bf R}_2} \sum_{\mu,\nu} \left[ t_{\mu\nu}^{r,t}\,
  d_{i,\mu}^{\dagger}\, p_{j,\nu}^{t} + t_{\mu\nu}^{r,b}\,
  d_{i,\mu}^{\dagger}\, p_{j,\nu}^{b} + {\rm H.c.}  \right]
\nonumber\\ & & +\sum_{{\bf r}_i} \sum_{{\bf r}_j={\bf r}_i-{\bf
    R}_{2}} \sum_{\mu,\nu} \left[ t_{\mu\nu}^{l,t}\,
  d_{i,\mu}^{\dagger}\, p_{j,\nu}^{t} + t_{\mu\nu}^{l,b}\,
  d_{i,\mu}^{\dagger}\, p_{j,\nu}^{b} + {\rm H.c.}  \right] \nonumber
\\ & & +\sum_{{\bf r}_i} \sum_{\nu,\nu'} \left[ s_{\nu\nu'} \left(
  p_{i,\nu}^{t} \right)^{\dagger} p_{i,\nu'}^{b} + {\rm H.c.}  \right]
\nonumber \\ & & + \sum_{\langle ij\rangle} \sum_{\mu,\mu'} \left[
  v_{\mu\mu'}^{ij}\, d_{i,\mu}^{\dagger}\, d_{j,\mu'} + {\rm H.c.}
  \right] \nonumber \\ & & + \sum_{\langle ij\rangle} \sum_{\nu,\nu'}
\left\{ u_{\nu\nu'}^{ij} \left[ \left( p_{i,\nu}^{t} \right)^{\dagger}
  p_{j,\nu'}^{t} + \left( p_{i,\nu}^{b} \right)^{\dagger}
  p_{j,\nu'}^{b} \right] + {\rm H.c.}  \right\},
\end{eqnarray}
where $\langle ij\rangle$ denotes a sum over pairs of
nearest-neighbour cells. The operators $d_{i,\mu}$
($d_{i,\mu}^{\dagger}$) annihilate (create) an electron on the Mo in
the unit cell $i$ in the orbital $\mu$. Similarly, the operators
$p_{i,\nu}^{b}$ [$(p_{i,\nu}^{b})^{\dagger}$] and $p_{i,\nu}^{t}$
[$(p_{i,\nu}^{t})^{\dagger}$] annihilate (create) electrons at the
bottom $b$ and top $t$ S sites of the unit cell $i$, respectively. We
assume that the top and bottom S layers are symmetric ($z$ inversion
symmetry).

We use the basis set defined in Sec.~\ref{sec:3model} to express the
on-site energies and the hopping integrals. The on-site energies are
given by
\begin{equation}
\varepsilon_{\mu}^{\rm{Mo}} = \langle {\bf r}_i; d_{\mu} |\mathcal{H}|
           {\bf r}_i; d_{\mu} \rangle
\end{equation}
and
\begin{equation}
\varepsilon_{\nu}^{{S}} = \langle {\bf r}_i + {\bm \delta}_{1\pm};
p_{\nu} |\mathcal{H}| {\bf r}_i + {\bm \delta}_{1\pm}; p_{\nu}
\rangle.
\end{equation}
The hopping matrix elements between Mo and S orbitals are
\begin{eqnarray}
t_{\mu\nu}^{t} & = & \langle {\bf r}_i; d_{\mu} |\mathcal{H}| {\bf
  r}_i + {\bm \delta}_{1+}; p_{\nu} \rangle, \\ t_{\mu\nu}^{b} & = &
\langle {\bf r}_i; d_{\mu} |\mathcal{H}| {\bf r}_i + {\bm
  \delta}_{1-}; p_{\nu} \rangle , \\ t_{\mu\nu}^{r,t} & = & \langle
        {\bf r}_i; d_{\mu} |\mathcal{H}| {\bf r}_i + {\bf R}_1 - {\bf
          R}_2 + {\bm \delta}_{1+}; p_{\nu} \rangle,
        \\ t_{\mu\nu}^{r,b} & = & \langle {\bf r}_i; d_{\mu}
        |\mathcal{H}| {\bf r}_i + {\bf R}_1 - {\bf R}_2 + {\bm
          \delta}_{1-}; p_{\nu} \rangle, \\ t_{\mu\nu}^{l,t} & = &
        \langle {\bf r}_i; d_{\mu} |\mathcal{H}| {\bf r}_i - {\bf R}_2
        + {\bm \delta}_{1+}; p_{\nu} \rangle, \\ t_{\mu\nu}^{l,b} & =
        & \langle {\bf r}_i; d_{\mu} |\mathcal{H}| {\bf r}_i - {\bf
          R}_2 + {\bm \delta}_{1-}; p_{\nu} \rangle.
\end{eqnarray}
The hopping matrix elements between top and bottom S orbitals read
\begin{eqnarray}
 s_{\nu\nu'} & = & \langle {\bf r}_i + {\bm \delta}_{1+};
p_{\mu} |\mathcal{H}| {\bf r}_i + {\bm \delta}_{1-}; p_{\nu'} \rangle,
\end{eqnarray}
while the nearest-neighbor Mo-Mo hopping integrals are
\begin{eqnarray}
v_{\mu\mu'}^{{E}} & = & \langle{\bf r}_i; d_{\mu} |\mathcal{H}| {\bf
  r}_i + {\bf R}_{1}; d_{\mu'} \rangle, \\ v_{\mu\mu'}^{\rm{NE}} & = &
\langle{\bf r}_i; d_{\mu} |\mathcal{H}| {\bf r}_i + {\bf R}_{2};
d_{\mu'} \rangle, \\ v_{\mu\mu'}^{\rm{NW}} & = & \langle{\bf r}_i;
d_{\mu} |\mathcal{H} |{\bf r}_i + {\bf R}_{2}-{\bf R}_{1}; d_{\mu'}
\rangle, \\ v_{\mu\mu'}^{{W}} & = &\langle{\bf r}_i; d_{\mu}
|\mathcal{H} |{\bf r}_i - {\bf R}_{1}; d_{\mu'} \rangle,
\\ v_{\mu\mu'}^{\rm{SW}} & = & \langle{\bf r}_i; d_{\mu} |\mathcal{H}|
   {\bf r}_i - {\bf R}_{2}; d_{\mu'} \rangle, \\ v_{\mu\mu'}^{\rm{SE}}
   & = & \langle{\bf r}_i; d_{\mu} |\mathcal{H}| {\bf r}_i - {\bf
     R}_{2} + {\bf R}_{1}; d_{\mu'} \rangle, \\
\end{eqnarray}
and the S--S next-nearest-neighbor hopping matrix elements read
\begin{eqnarray}
u_{\nu\nu'}^{\rm{E}} & = & \langle{\bf r}_i + {\bm \delta}_{1\pm};
p_{\nu} |\mathcal{H}| {\bf r}_i + {\bf R}_{1} + {\bm \delta}_{1\pm};
p_{\nu'} \rangle, \\ u_{\nu\nu'}^{\rm{NE}} & = & \langle{\bf r}_i +
{\bm \delta}_{1\pm}; p_{\nu} |\mathcal{H}| {\bf r}_i + {\bf R}_{2} +
{\bm \delta}_{1\pm}; p_{\nu'} \rangle, \\ u_{\nu\nu'}^{\rm{NW}} & = &
\langle{\bf r}_i + {\bm \delta}_{1\pm}; p_{\nu}|\mathcal{H}| {\bf r}_i
+ {\bf R}_{2} - {\bf R}_{1} + {\bm \delta}_{1\pm}; p_{\nu'} \rangle,
\\ u_{\nu\nu'}^{{W}} & = & \langle{\bf r}_i + {\bm \delta}_{1\pm};
p_{\nu} |\mathcal{H}| {\bf r}_i - {\bf R}_{1} + {\bm \delta}_{1\pm};
p_{\nu'} \rangle, \\ u_{\nu\nu'}^{\rm{SW}} & = & \langle{\bf r}_i +
{\bm \delta}_{1\pm}; p_{\nu} |\mathcal{H}| {\bf r}_i- {\bf R}_{2} +
{\bm \delta}_{1\pm}; p_{\nu'} \rangle, \\ u_{\nu\nu'}^{\rm{SE}} & = &
\langle{\bf r}_i + {\bm \delta}_{1\pm}; p_{\nu} |\mathcal{H}| {\bf
  r}_i- {\bf R}_{2} + {\bf R}_{1} + {\bm \delta}_{1\pm}; p_{\nu'}
\rangle.
\end{eqnarray}

\begin{figure}[h]
\centering
\includegraphics[width=0.7\columnwidth]{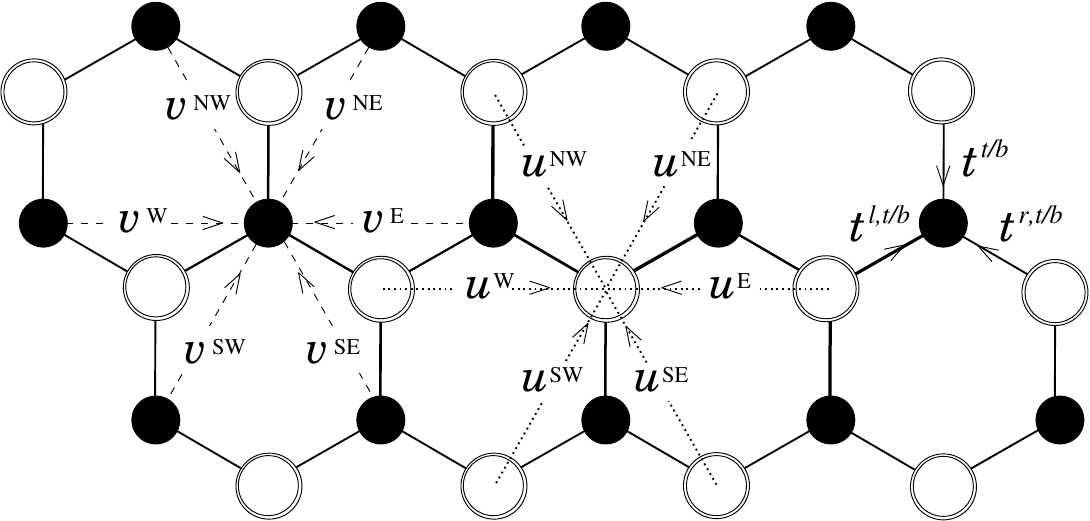} 
\caption{Scheme of the hopping amplitudes. Solid black circles
  represent Mo atoms, while empty circles represent the S atoms at the
  top and bottom layers.}
\label{nn_hops}
\end{figure}

Notice that $\mathcal{H}$ contains Mo--S and S--S nearest-neighbour
(same unit cell) hopping amplitudes and Mo--Mo and S--S
next-to-nearest-neighbour hopping amplitudes (adjacent cells). For the
latter, each Mo and each S has six next-to-nearest neighbours. For the
former, each Mo has six S nearest neighbours. Overall, there is a
total of 25 hopping amplitudes within the unit cell and between the
unit cell and the adjacent cells. The hopping amplitudes are indicated
in Fig.~\ref{nn_hops}.

%---------------------------------------------------------
%----------------------------------------------------------

The following associations are made for the on-site energies of Mo
atoms: $\varepsilon_{0}^{\rm{Mo}} \equiv \Delta_{0}$,
$\varepsilon_{1}^{\rm{Mo}} = \varepsilon_{2}^{\rm{Mo}} \equiv
\Delta_{2}$, $ \varepsilon_{3}^{\rm{Mo}} = \varepsilon_{4}^{\rm{Mo}}
\equiv \Delta_{1}$. For on-site energies of the S atoms we define
$\varepsilon_{1}^{{S}} = \varepsilon_{2}^{{S}} \equiv \Delta_{p}$ and
$\varepsilon_{3}^{{S}} \equiv \Delta_{z}$.

The hopping amplitudes can be written in terms of SK integrals. We can
also incorporate the $x$ and $z$ reflection symmetries, when
applicable, to reduce the number of terms. Here we provide expressions
for the more relevant hopping amplitudes in terms of seven SK
integrals. This allows us to substantially reduce the number of
fitting parameters of the model. Below we list amplitudes that are not
identically zero.

\begin{itemize}
\item Mo--S (Here we present the expressions for the $t^t$ hopping
  matrix elements. The $t^b$ ones follow similar expressions, with
  $\theta_b \leftrightarrow -\theta_b$.)
\begin{equation}
\hskip-3.5cm
t_{02}^{t}  =  \cos\theta_{b} \left( \sin^{2} \theta_{b} -
\frac{1}{2} \cos^{2} \theta_{b} \right) V_{pd\sigma} - \sqrt{3} \cos
\theta_{b} \sin^{2} \theta_{b}V_{pd\pi},
\end{equation}
\begin{equation}
\hskip-3.5cm
t_{03}^{t}  =  \sin\theta_{b} \left( \sin^{2} \theta_{b} - 
\frac{1}{2} \cos^{2} \theta_{b} \right) V_{pd\sigma} + 
\sqrt{3} \sin \theta_{b} \cos^{2} \theta_{b} V_{pd\pi},
\end{equation}
\begin{equation}
\hskip-3.5cm
t_{12}^{t}  =  - \frac{\sqrt{3}}{2} \cos^{3} \theta_{b} V_{pd\sigma} - 
\cos \theta_{b} \sin^{2} \theta_{b} V_{pd\pi},
\end{equation}
\begin{equation}
\hskip-3.5cm
t_{13}^{t}  =  -\sin \theta_{b} \cos^{2} \theta_{b}
\left( \frac{\sqrt{3}}{2} V_{pd\sigma} - V_{pd\pi} \right),
\end{equation}
\begin{equation}
\hskip-3.5cm
 t_{21}^{t} = \cos \theta_{b} V_{pd\pi},
\end{equation}
\begin{equation}
\hskip-3.5cm
 t_{31}^{t} = \sin\theta_{b} V_{pd\pi},
\end{equation}
\begin{equation} 
\hskip-3.5cm
  t_{42}^{t} = \sqrt{3} \cos^{2} \theta_{b}
\sin \theta_{b} V_{pd\sigma} + \sin \theta_{b} (1 - 2 \cos^{2}
\theta_{b}) V_{pd\pi},
\end{equation}
\begin{equation}
\hskip-3.5cm
t_{43}^{t} = \sqrt{3} \sin^{2} \theta_{b}
\cos \theta_{b} V_{pd\sigma} + \cos \theta_{b} ( 1 -2 \sin^{2}
\theta_{b}) V_{pd\pi},
\end{equation}
\begin{equation}
\hskip-3.5cm
t_{01}^{l,t} = - \frac{\sqrt{3}}{2} \cos\theta_{b} \left(
\sin^{2}\theta_{b} - \frac{1}{2} \cos^{2}\theta_{b} \right)
V_{pd\sigma} + \frac{3}{2} \cos\theta_{b} \sin^{2}\theta_{b} V_{pd\pi},
\end{equation}
\begin{equation} 
\hskip-3.5cm
t_{02}^{l,t} = - \frac{1}{2} \cos\theta_{b} \left(
\sin^{2}\theta_{b} - \frac{1}{2}\cos^{2}\theta_{b} \right)
V_{pd\sigma} + \frac{\sqrt{3}}{2} \cos\theta_{b} \sin^{2}\theta_{b}
V_{pd\pi},
\end{equation}
\begin{equation}
\hskip-3.5cm
t_{03}^{l,t} = \sin\theta_{b} \left(
\sin^{2}\theta_{b} - \frac{1}{2} \cos^{2}\theta_{b} \right)
V_{pd\sigma} + \sqrt{3} \sin\theta_{b} \cos^{2}\theta_{b} V_{pd\pi},
\end{equation}
\begin{equation}
\hskip-3.5cm
t_{11}^{l,t} = - \frac{3}{8} \cos^{3}\theta_{b} V_{pd\sigma}-
\frac{\sqrt{3}}{2} \cos\theta_{b} \left( 1 - \frac{1}{2}
\cos^{2}\theta_{b} \right) V_{pd\pi},
\end{equation}
\begin{equation}
\hskip-3.5cm
t_{12}^{l,t} = -
\frac{\sqrt{3}}{8} \cos^{3}\theta_{b} V_{pd\sigma} + \frac{1}{2}
\cos\theta_{b} \left( 1 + \frac{1}{2} \cos^{2}\theta_{b} \right)V_{pd\pi}, 
\end{equation}
\begin{equation}
\hskip-3.5cm
t_{13}^{l,t} = \frac{1}{2} \sin\theta_{b}
\cos^{2}\theta_{b} \left( \frac{\sqrt{3}}{2} V_{pd\sigma} - V_{pd\pi},
\right) 
\end{equation}
\begin{equation}
\hskip-3.5cm
t_{21}^{l,t} = - \frac{3\sqrt{3}}{8} \cos^{3}\theta_{b}
V_{pd\sigma} - \frac{1}{2} \cos\theta_{b} \left( 1 - \frac{3}{2}
\cos^{2}\theta_{b} \right) V_{pd\pi}, 
\end{equation}
\begin{equation}
\hskip-3.5cm
 t_{22}^{l,t} = -
\frac{3}{8} \cos^{3} \theta_{b} V_{pd\sigma} -
\frac{\sqrt{3}}{2}\cos\theta_{b} \left( 1 - \frac{1}{2}
\cos^{2}\theta_{b} \right) V_{pd\pi}, 
\end{equation}
\begin{equation}
\hskip-3.5cm
 t_{23}^{l,t} =
\frac{\sqrt{3}}{4} \sin\theta_{b} \cos^{2}\theta_{b} \left( \sqrt{3}
V_{pd\sigma} - 2V_{pd\pi} \right),
\end{equation}
\begin{equation}
\hskip-3.5cm
t_{31}^{l,t} =
\frac{3\sqrt{3}}{4} \cos^{2}\theta_{b} \sin\theta_{b} V_{pd\sigma} +
\sin\theta_{b} \left( 1 - \frac{3}{2}\cos^{2} \theta_{b} \right)
V_{pd\pi}, 
\end{equation}
\begin{equation}
\hskip-3.5cm
t_{32}^{l,t} = \frac{\sqrt{3}}{4} \sin\theta_{b}
\cos^{2}\theta_{b} \left( \sqrt{3} V_{pd\sigma} - 2 V_{pd\pi} \right),
\end{equation}
\begin{equation}
\hskip-3.5cm
 t_{33}^{l,t} = - \frac{3}{2} \sin^{2}\theta_{b} \cos\theta_{b}
V_{pd\sigma} - \frac{\sqrt{3}}{2} \cos\theta_{b} \left( 1 - 2
\sin^{2}\theta_{b} \right) V_{pd\pi},
\end{equation}
\begin{equation}
\hskip-3.5cm
 t_{41}^{l,t} = 
\frac{\sqrt{3}}{4} \sin\theta_{b} \cos^{2}\theta_{b} \left( \sqrt{3}
V_{pd\sigma} - 2V_{pd\pi} \right), 
\end{equation}
\begin{equation}
\hskip-3.5cm
 t_{42}^{l,t} = 
\frac{\sqrt{3}}{4} \cos^{2}\theta_{b} \sin\theta_{b} V_{pd\sigma} +
\sin\theta_{b} \left( 1 - \frac{1}{2}\cos^{2}\theta_{b} \right)
V_{pd\pi},
\end{equation}
\begin{equation}
\hskip-3.5cm
 t_{43}^{l,t} = - \frac{\sqrt{3}}{2}
\sin^{2}\theta_{b} \cos\theta_{b} V_{pd\sigma} - \frac{1}{2}
\cos\theta_{b} \left( 1 - 2 \sin^{2}\theta_{b} \right) V_{pd\pi}.
\end{equation}
\item Mo--Mo ($v^W$, $v^{NW}$, $v^{SW}$ and $s^{SE}$ can be 
obtained from $v^E$ and $v^{NE}$ by symmetry.)
\begin{equation}
\hskip-3.5cm
v_{00}^{{E}} = \frac{1}{4} V_{dd\sigma}+\frac{3}{4} V_{dd\delta},
\end{equation}
\begin{equation}
\hskip-3.5cm 
v_{01}^{{E}} = -\frac{\sqrt{3}}{4} V_{dd\sigma} +
\frac{\sqrt{3}}{4} V_{dd\delta},
\end{equation}
\begin{equation}
\hskip-3.5cm 
v_{11}^{{E}} = \frac{3}{4} V_{dd\sigma} + \frac{1}{4} V_{dd\delta} 
\end{equation}
\begin{equation}
\hskip-3.5cm
v_{22}^{{E}} =  V_{dd\pi},
\end{equation}
\begin{equation}
\hskip-3.5cm 
v_{33}^{{E}}  =  V_{dd\pi},
\end{equation}
\begin{equation}
\hskip-3.5cm 
v_{44}^{{E}}  =  V_{dd\delta},
\end{equation}
\begin{equation}
\hskip-3.5cm 
v_{02}^{\rm{NE}} = -\frac{3}{8} V_{dd\sigma} +
\frac{3}{8}V_{dd\delta},
\end{equation}
\begin{equation}
\hskip-3.5cm 
v_{12}^{\rm{NE}} = - \frac{3\sqrt{3}}{16}V_{dd\sigma} +
\frac{\sqrt{3}}{4} V_{dd\pi} - \frac{\sqrt{3}}{16} V_{dd\delta},
\end{equation}
\begin{equation}
\hskip-3.5cm 
v_{34}^{\rm{NE}} = \frac{\sqrt{3}}{4} V_{dd\pi} -\frac{\sqrt{3}}{4}
V_{dd\delta}.
\end{equation}
\item S--S ($u^{W}$, $u^{NW}$, $u^{SW}$, and $u^{SE}$ can be obtained
  from $u^E$ and $u^{NE}$ by symmetry.)
\begin{equation}
\hskip-3.5cm
u_{11}^{{E}}  =  V_{pp\sigma},
\end{equation}
\begin{equation}
\hskip-3.5cm
u_{22}^{{E}}  =  V_{pp\pi},
\end{equation}
\begin{equation}
\hskip-3.5cm
u_{33}^{{E}}  =  V_{pp\pi},
\end{equation}
\begin{equation}
\hskip-3.5cm
u_{12}^{\rm{NE}}  =  \frac{\sqrt{3}}{4}(V_{pp\sigma}-V_{pp\pi}).
\end{equation}
\begin{equation}
\hskip-3.5cm
s_{11}  =  V_{pp\pi},
\end{equation}
\begin{equation}
\hskip-3.5cm
s_{22}  =  V_{pp\pi},
\end{equation}
\begin{equation}
\hskip-3.5cm
s_{33}  =  V_{pp\sigma}.
\end{equation}
 
\end{itemize}

A general yet compact expression for all amplitudes is given by the
matrices
\begin{equation}
\hskip-3.5cm
t^t = \left( \begin{array}{ccc} 0 & t^t_{02} & t^t_{03} \\ 0 &
  t^t_{12} & t^t_{13} \\ t^t_{21} & 0 & 0 \\ t^t_{31} & 0 & 0 \\ 0 &
  t^t_{42} & t^t_{43}
\end{array} \right),
\end{equation}
\begin{equation}
\hskip-3.5cm
t^b = \left( \begin{array}{ccc} 0 & t^t_{02} & - t_{03}^t \\ 0 &
  t_{12}^t & - t_{13}^t \\ t_{21}^t & 0 & 0 \\ - t_{31}^t & 0 & 0 \\ 0
  & - t_{42}^t & t_{43}^t
\end{array} \right),
\end{equation}
\begin{equation}
\hskip-3.5cm
t^{l,t} = \left( \begin{array}{ccc} t_{01}^{l,t} & t^{l,t}_{02}
  & t^{l,t}_{03} \\ t_{11}^{l,t} & t^{l,t}_{12} & t^{l,t}_{13}
  \\ t^{l,t}_{21} & t_{22}^{l,t} & t_{23}^{l,t} \\ t^{l,t}_{31} &
  t_{32}^{l,t} & t_{33}^{l,t} \\ t_{41}^{l,t} & t^{l,t}_{42} &
  t^{l,t}_{43}
\end{array} \right),
\end{equation}
\begin{equation}
\hskip-3.5cm
t^{l,b} = \left( \begin{array}{ccc} t_{01}^{l,t} & t^{l,t}_{02}
  & - t^{l,t}_{03} \\ t_{11}^{l,t} & t^{l,t}_{12} & - t^{l,t}_{13}
  \\ t^{l,t}_{21} & t_{22}^{l,t} & - t_{23}^{l,t} \\ - t^{l,t}_{31} &
  - t_{32}^{l,t} & t_{33}^{l,t} \\ - t_{41}^{l,t} & - t^{l,t}_{42} &
  t^{l,t}_{43}
\end{array} \right),
\end{equation}
\begin{equation}
\hskip-3.5cm
  t^{r,t} = \left( \begin{array}{ccc} - t_{01}^{l,t} &
      t^{l,t}_{02} & t^{l,t}_{03} \\ - t_{11}^{l,t} & t^{l,t}_{12} &
      t^{l,t}_{13} \\ t^{l,t}_{21} & - t_{22}^{l,t} & - t_{23}^{t {L}}
      \\ t^{l,t}_{31} & - t_{32}^{l,t} & - t_{33}^{l,t} \\ - t_{41}^{l,t}
      & t^{l,t}_{42} & t^{l,t}_{43}
\end{array} \right),
\end{equation}
\begin{equation}
\hskip-3.5cm
t^{r,b} = \left( \begin{array}{ccc} - t_{01}^{l,t} &
  t^{l,t}_{02} & - t^{l,t}_{03} \\ - t_{11}^{l,t} & t^{l,t}_{12} & -
  t^{l,t}_{13} \\ t^{l,t}_{21} & - t_{22}^{l,t} & t_{23}^{l,t} \\ -
  t^{l,t}_{31} & t_{32}^{l,t} & - t_{33}^{l,t} \\ t_{41}^{l,t} & -
  t^{l,t}_{42} & t^{l,t}_{43}
\end{array} \right),
\end{equation}
\begin{equation}
\hskip-3.5cm
v^{{E}} = \left( \begin{array}{ccccc} v^{{E}}_{00} & v^{{E}}_{01} &
  0 & 0 & 0 \\ v^{{E}}_{01} & v^E_{11} & 0 & 0 &
  0 \\ 0 & 0 & v^{{E}}_{22} & 0 & 0 \\ 0 & 0 & 0
  & v^{{E}}_{33} & 0 \\ 0 & 0 & 0 & 0 &
  v^{{E}}_{44}
\end{array} \right),
\end{equation}
\begin{equation}
\hskip-3.5cm
  v^{{W}} = \left( \begin{array}{ccccc} v^{{E}}_{00} & v^{{E}}_{01} &
  0 & 0 & 0 \\ v^{{E}}_{01} & v^E_{11} & 0 & 0 &
  0 \\ 0 & 0 & v^{{E}}_{22} & 0 & 0 \\ 0 & 0 & 0
  & v^{{E}}_{33} & 0 \\ 0 & 0 & 0 & 0 &
  v^{{E}}_{44}
\end{array} \right),
\end{equation}
\begin{equation}
\hskip-3.5cm
  v^{{NE}} = \left( \begin{array}{ccccc} v_{00}^E & - \frac{1}{2}
      v_{01}^E & v_{02}^{{NE}} & 0 & 0 \\ - \frac{1}{2} v_{01}^E &
      \frac{1}{4} v_{11}^{{E}} + \frac{3}{4} v_{22}^{{E}} & v_{12}^{{NE}}
      & 0 & 0 \\ v_{02}^{{NE}} & v_{12}^{{NE}} & \frac{1}{4} v_{22}^{{E}}
      + \frac{3}{4} v_{11}^{{E}} & 0 & 0 \\ 0 & 0 & 0 & \frac{1}{4}
      v_{33}^{{E}} + \frac{3}{4} v_{44}^{{E}} & v_{34}^{{NE}} \\ 0 & 0 & 0
      & v_{34}^{{NE}} & \frac{1}{4} v_{44}^{{E}} + \frac{3}{4}
      v_{33}^{{E}}
\end{array} \right),
\end{equation}
\begin{equation}
\hskip-3.5cm
v^{{NW}} = \left( \begin{array}{ccccc} v_{00}^E & - \frac{1}{2}
  v_{01}^E & - v_{02}^{{NE}} & 0 & 0 \\ - \frac{1}{2} v_{01}^E &
  \frac{1}{4} v_{11}^{{E}} + \frac{3}{4} v_{22}^{{E}} & -
  v_{12}^{{NE}} & 0 & 0 \\ - v_{02}^{{NE}} & - v_{12}^{{NE}} &
  \frac{1}{4} v_{22}^{{E}} + \frac{3}{4} v_{11}^{{E}} & 0 & 0 \\ 0 & 0
  & 0 & \frac{1}{4} v_{33}^{{E}} + \frac{3}{4} v_{44}^{{E}} & -
  v_{34}^{{NE}} \\ 0 & 0 & 0 & - v_{34}^{{NE}} & \frac{1}{4}
  v_{44}^{{E}} + \frac{3}{4} v_{33}^{{E}}
\end{array} \right),
\end{equation}
\begin{equation}
\hskip-3.5cm
u^{{E}} = \left( \begin{array}{ccc} u^{{E}}_{11} & 0 & 0 \\ 0 &
  u^{{E}}_{22} & 0 \\ 0 & 0 & u^{{E}}_{33}
\end{array} \right),
\end{equation}
\begin{equation}
\hskip-3.5cm
u^{{NE}} = \left( \begin{array}{ccc} \frac{1}{4} u_{11}^{{E}} +
  \frac{3}{4} u_{22}^{{E}} & u_{12}^{{NE}} & 0 \\ u_{12}^{{NE}} &
  \frac{1}{4} u_{22}^{{E}} + \frac{3}{4} u_{11}^{{E}} & 0 \\ 0 & 0 &
  u_{33}^{{E}}
\end{array} \right),
\end{equation}
and
\begin{equation}
\hskip-3.5cm
u^{{NW}} = \left( \begin{array}{ccc} \frac{1}{4} u_{11}^{{E}} +
  \frac{3}{4} u_{22}^{{E}} & - u_{12}^{{NE}} & 0 \\ - u_{12}^{{NE}} &
  \frac{1}{4} u_{22}^{{E}} + \frac{3}{4} u_{11}^{{E}} & 0 \\ 0 & 0 &
  u_{33}^{{E}}
\end{array} \right).
\end{equation}
%

%%%%%%%%%%%%%%%%%%%%%%%%%%%%%%%%%%%%%%%%%%%%
\section{Tight-binding energy bands}
\label{subsec:Appendix B}

To find the energy bands we need to solve the the
eigenvalue/eigenvector problem in the Bloch momentum representation,
\begin{eqnarray}
\mathcal{H}|{\bf k}\rangle = E({\bf k}) |{\bf k}\rangle,
\label{eq:schrondigenr}
\end{eqnarray}
where the Bloch vector is given by 
\begin{eqnarray}
\hskip-1.8cm
|{\bf k}\rangle & = & \sum_{{\bf r}_i} e^{i{\bf k}\cdot{\bf r}_i}
\left[ \sum_{\mu=0}^{4} \alpha_{{\bf k},\mu}| {\bf r}_i; d_{\mu}
  \rangle + \sum_{\mu=1}^{3} \left( \beta_{{\bf k},\mu}| {\bf r}_i +
          {\bm \delta}_{1-}; p_{\mu} \rangle + \tau_{{\bf k},\mu}|
          {\bf r}_i + {\bm \delta}_{1+}; p_{\mu} \rangle \right)
          \right].
\label{eq:Bloch}
\end{eqnarray}
Resorting to the orthogonal bais, we can rewrite the
eigenvalue/eigenvector problem as a system of linear coupled
equations, 
\begin{eqnarray}
\hskip-1.8cm
\left[ E(\bf{k}) - \varepsilon_{{\mu}}^{\rm{Mo}} \right]
\alpha_{\bf{k},{\mu}} & = & \sum_{\nu} \left[ \left( t_{{\mu}\nu}^{t}
  + z_{2}^{\ast}\, t_{{\mu}\nu}^{t{L}} + z_{2}^{\ast}\, z_{1}
  t_{{\mu}\nu}^{r,t} \right) \tau_{\bf{k},\nu} 
% \right.   \nonumber \\ & & \left. 
+ \left( t_{{\mu}\nu}^{b} + z_{2}^{\ast}\,
  t_{{\mu}\nu}^{l,b} + z_{2}^{\ast}\, z_{1}\, t_{{\mu}\nu}^{r,b}
  \right) \beta_{\bf{k},\nu} \right] \nonumber \\  &&  + 2\sum_{{\mu}'}
\left( c_{1}\, v_{{\mu}{\mu}'}^{{E}} + c_{2}\,
v_{{\mu}{\mu}'}^{\rm{NE}} + c_{12}\, v_{{\mu}{\mu}'}^{\rm{NW}} \right)
\alpha_{\bf{k},{\mu}'},
\end{eqnarray}
\begin{eqnarray}
\hskip-1.8cm
  \left[ E({\bf k}) - \varepsilon_{\nu}^{{\rm S}} \right] \beta_{\bf{k},\nu}
  & = & \sum_{{\mu}} \left( t_{{\mu}\nu}^{b} + z_{2}  t_{{\mu}\nu}^{l,b} 
    + z_{2}\, z_{1}^{\ast} t_{{\mu}\nu}^{r,b} \right) \alpha_{\bf{k},{\mu}}
  \nonumber \\ & & + \sum_{\nu'} \left[ \left( c_{1}\, u_{\nu\nu'}^{{E}}
      + c_{2}\, u_{\nu\nu'}^{\rm{NE}} + c_{12}\, u_{\nu\nu'}^{\rm{NW}} 
    \right) \beta_{\bf{k},\nu'} + s_{\nu\nu'}\, \tau_{\bf{k},\nu'}
  \right],
\end{eqnarray}
and
\begin{eqnarray}
\hskip-1.8cm
  \left[E({\bf k}) - \varepsilon_{\nu}^{{\rm S}} \right] \tau_{\bf{k},\nu} &
  = & \sum_{{\mu}} \left( t_{{\mu}\nu}^t + z_{2}
    t_{{\mu}\nu}^{t{L}} + z_{2}\, z_{1}^{\ast} t_{{\mu}\nu}^{t{R}} 
  \right) \alpha_{\bf{k},{\mu}}
  \nonumber \\ & & + \sum_{\nu'} \left[ \left( c_{1}\, u_{\nu\nu'}^{{E}}
      + c_{2}\, u_{\nu\nu'}^{\rm{NE}} + c_{12}\, u_{\nu\nu'}^{\rm{NW}}
    \right) \tau_{\bf{k},\nu'} + s_{\nu\nu'} \beta_{\bf{k},\nu'}
  \right],
\end{eqnarray}
where $z_{1} = e^{i{\bf k}\cdot{\bf R}_{1}} = e^{iak_{x}}$, $z_{2} =
e^{i{\bf k}\cdot{\bf R}_{2}} = e^{iak_{x}/2} e^{i\sqrt{3}ak_{y}/2}$,
\begin{equation}
\label{eq:c1}
c_{1} \equiv \frac{z_{1} + z_{1}^{\ast}}{2} = \cos({\bf k} \cdot {\bf
  R}_{1}) = \cos(k_{x}a),
\end{equation}
\begin{equation}
\label{eq:c2}
c_{2} \equiv \frac{z_{2} + z_{2}^{\ast}}{2} = \cos({\bf k} \cdot {\bf
  R}_{2}) = \cos \left( k_{x} a/2 + k_{y} \sqrt{3} a/2 \right),
\end{equation}
and 
\begin{eqnarray}
\label{eq:c12}
c_{12} \equiv \frac{z_{1}\, z_{2}^{\ast} + z_{1}^{\ast}\, z_{2}}{2} =
\cos({\bf k} \cdot {\bf R}_{1} - {\bf k} \cdot {\bf R}_{2}) = \cos
\left( k_{x} a/2 - k_{y} \sqrt{3} a/2 \right).
\end{eqnarray}

In matrix form, we have 
\begin{eqnarray}
  \left(\begin{array}{ccc} h^{\rm{Mo}} + V & T^{t} &
      T^{b} \\ \left( T^{t} \right)^T & h^{{S}}
      + U & s \\ \left( T^{b} \right)^T &
      s & h^{{S}} + U
\end{array} \right) \left( \begin{array}{c}
\alpha \\ \tau \\ \beta
\end{array} \right) = E\left( \begin{array}{c}
\alpha \\ \tau \\ \beta
\end{array} \right),
\label{eq:eigen_matrix}
\end{eqnarray}
where
\begin{equation}
\hskip-2.0cm
  h^{\rm{Mo}} = \left( \begin{array}{ccccc}
      \Delta_{0} & 0 & 0 & 0 & 0\\
      0 & \Delta_{2} & 0 & 0 & 0\\
      0 & 0 & \Delta_{2} & 0 & 0\\
      0 & 0 & 0 & \Delta_{1} & 0\\
      0 & 0 & 0 & 0 & \Delta_{1}
\end{array} \right),
\end{equation}
\begin{equation}
\hskip-2.0cm
  h^{{S}} = \left( \begin{array}{ccc} \Delta_{p} & 0 & 0\\ 0 &
      \Delta_{p} & 0\\ 0 & 0 & \Delta_{z}
\end{array} \right),
\end{equation}
\begin{equation}
\hskip-2.cm
  T^{t} = t^{t} + z_{2}^{\ast} \left( t^{l,t} +
    z_{1}\, t^{r,t} \right),
\label{eq:Tt}
\end{equation}
\begin{equation}
\hskip-2.0cm
  T^{b} = t^{b} + z_{2}^{\ast} \left( t^{l,b} +
    z_{1}\, t^{r,b} \right),
\end{equation}
\begin{equation}
\hskip-2.cm
  V  =  2 \left( c_{1}\, v^{{E}} + c_{2}\,
    v^{\rm{NE}} + c_{12}\, v^{\rm{NW}} \right),
\end{equation}
\mbox{and}
\begin{equation}
\hskip-2.0cm
  U  =  2 \left( c_{1}\, u^{{E}} + c_{2}\,
    u^{\rm{NE}} + c_{12}\, u^{\rm{NW}} \right).
\end{equation}

This formulation suffices for a numerical evaluation of the
bands. However, in order to obtain analytical expression for the gap
and other features of the band structure at the symmetry points, it is
necessary to reduce the size of the matrices to be diagonalized. This
can be done by exploring underlying symmetries in the equations.

Let us define the symmetric and anti-symmetric components
\begin{equation}
  \theta_{{\bf k},\nu} = \frac{1}{\sqrt{2}}(\tau_{{\bf k},\nu} +
  \beta_{{\bf k},\nu})
\end{equation}
and
\begin{equation}
  \phi_{{\bf k},\nu} = \frac{1}{\sqrt{2}}(\tau_{{\bf k},\nu} -
  \beta_{{\bf k},\nu}).
\end{equation}
In terms of these components, the eigenproblem takes the form
\begin{eqnarray}
  \left( \begin{array}{ccc}
      h^{\rm{Mo}} + V & T^{E} & T^{O} \\
      T^{E\dagger} & h_{}^{{S}} + U + s & 0 \\
      T^{O\dagger} & 0 & h^{{S}} + U - s
\end{array} \right) \left( \begin{array}{c}
  \alpha \\
  \theta \\
  \phi
\end{array} \right) = E \left( \begin{array}{c}
  \alpha \\
  \theta \\
  \phi
\end{array} \right),
\end{eqnarray}
where we introduced new hopping matrices
\begin{equation}
T^{E} = \frac{1}{\sqrt{2}} \left( T^t + T^b \right) 
\end{equation}
and
\begin{equation}
T^{O} = \frac{1}{\sqrt{2}} \left( T^t - T^b \right).
\end{equation}

We can further simplify the eigenproblem by rearranging amplitudes in
the eigenvector, going from
\begin{equation}
  \psi^T = \left(
      \alpha_{0}, \alpha_{1}, \alpha_{2}, \alpha_{3}, \alpha_{4}, 
      \theta_{1}, \theta_{2}, \theta_{3}, \phi_{1}, \phi_{2}, \phi_{3} 
     \right).
\end{equation}
to
\begin{equation}
  \tilde{\psi}^T = \left(
        \alpha_{0}, \alpha_{1}, \alpha_{2}, \theta_{1}, \theta_{2}, 
        \phi_{3}, \alpha_{3}, \alpha_{4}, \phi_{1}, \phi_{2}, \theta_{3} 
     \right).
\end{equation}
This amounts to ordering the basis states such that the first six
components in the eigenvector are even ($E$) while the last five are
odd ($O$) with respect to $z$ inversion. As a result, the eigenproblem
can be recast in the decoupled form
\begin{equation}
  \left( \begin{array}{cc} H^{E} & 0 \\ 0 & H^{O} \end{array} 
  \right) \tilde{\psi} = E \tilde{\psi},
\end{equation}
where 
\begin{equation}
H^{E} = \left( \begin{array}{cc}
h^{\rm{Mo},E} & T^{E,O} \\
T^{E,O\dagger} & h^{S,E}
\end{array} \right)
\end{equation}
and
\begin{equation}
H^{O} = \left( \begin{array}{cc}
h^{\rm{Mo},O} & T^{O,E} \\
T^{O,E\dagger} & h^{S,O}
\end{array} \right).
\end{equation}
We have introduced the following matrices: 
\begin{equation}
  h^{\rm{Mo},E} = \left( \begin{array}{ccc}
      \Delta_{0} & 0 & 0 \\ 0 & \Delta_{2} & 0 \\ 0 & 0 & \Delta_{2}
\end{array} \right) + V^{E},
\end{equation}
\begin{equation}
  h^{\rm{Mo},O} = \left( \begin{array}{cc} \Delta_{1} & 0 \\ 0 & \Delta_{1}
\end{array} \right) + V^{O},
\end{equation}
\begin{equation}
  h^{S,O} = \left( \begin{array}{ccc} \Delta_{p} & 0 & 0 \\
      0 & \Delta_{p} & 0 \\ 0 & 0 & \Delta_{z}
\end{array} \right) + U^{E,O} + S^{O},
\end{equation}
\begin{equation}
  h^{S,E} = \left( \begin{array}{ccc} \Delta_{p} & 0 & 0 \\
      0 & \Delta_{p} & 0 \\ 0 & 0 & \Delta_{z}
\end{array} \right) + U^{E,O} + S^{E},
\end{equation}
\begin{equation}
U^{E,O} = U,
\end{equation}
\begin{equation}
  T^{E,O} = \left( \begin{array}{ccc}
      T_{01}^{E} & T_{02}^{E} & T_{03}^{O} \\
      T_{11}^{E} & T_{12}^{E} & T_{13}^{O} \\
      T_{21}^{E} & T_{22}^{E} & T_{23}^{O} \end{array} \right),
\end{equation}
\begin{equation}
  T^{O,E} = \left( \begin{array}{ccc}
      T_{31}^{O} & T_{32}^{O} & T_{33}^{E} \\
      T_{41}^{O} & T_{42}^{O} & T_{43}^{E}
\end{array} \right),
\end{equation}
\begin{equation}
  V^{E} = \left( \begin{array}{ccc}
      V_{00} & V_{01} & V_{02} \\
      V_{01} & V_{11} & V_{12} \\
      V_{02} & V_{12} & V_{22}
\end{array} \right),
\end{equation}
\begin{equation}
  V^{O} = \left( \begin{array}{cc}
      V_{33} & V_{34} \\
      V_{34} & V_{44}
\end{array} \right),
\end{equation}
\begin{equation}
  U^{E,O} = \left( \begin{array}{ccc}
      U_{11} & U_{12} & 0 \\
      U_{12} & U_{22} & 0 \\
      0 & 0 & U_{33}
\end{array} \right)
\end{equation}
\begin{equation}
  S^{O} = \left( \begin{array}{ccc}
      -s_{11} & 0 & 0 \\
      0 & -s_{22} & 0 \\
      0 & 0 & s_{33}
\end{array} \right),
\end{equation}
\mbox{and}
\begin{equation}
  S^{E} = \left( \begin{array}{ccc}
      s_{11} & 0 & 0 \\
      0 & s_{22} & 0 \\
      0 & 0 & -s_{33}
\end{array} \right).
\end{equation}
%

%%%%%%%%%%%%%%%%%%%%%%%%%%%%%%%%%%%%%%%%%%%%%%
\section{Expansion around symmetry points}
\label{subsec:Appendix C}

Expanding the hopping matrix elements around symmetry points in the
Brillouin zone allows to obtain analytical expressions for bands
energies and orbital composition.

%======================================================================
\subsection{$\Gamma$ point}

At the $\Gamma$ point, $k_{x}=k_{y}=0$, resulting in $z_{1}=z_{2}=1$
and $c_{1}=c_{2}=c_{12}=1$. Then,
\begin{equation}
  T_{\Gamma}^{E,O} = \sqrt{2} \left( \begin{array}{ccc}
      0 & 0 & t_{03}^{t} + 2\, t_{03}^{l,t} \\
      0 & t_{12}^{t} + , t_{12}^{l,t} & 0 \\
      t_{21}^{t} + 2\, t_{21}^{l,t} & 0 & 0
\end{array} \right),
\end{equation}
\begin{equation}
  T_{\Gamma}^{O,E} = \sqrt{2} \left( \begin{array}{ccc}
      t_{31}^{t} + 2\, t_{31}^{l,t} & 0 & 0 \\
      0 & t_{42}^{t} + 2\, t_{42}^{l,t} & 0
\end{array} \right),
\end{equation}
\begin{equation}
  V_{\Gamma}^{E} = 2 \left( \begin{array}{ccc}
      3v_{00}^{{E}} & 0 & 0 \\
      0 & \frac{3}{2} \left( v_{11}^{{E}} + v_{22}^{{E}} \right) & 0 \\
      0 & 0 & \frac{3}{2}\left(v_{11}^{{E}}+v_{22}^{{E}} \right)
\end{array} \right),
\end{equation}
\begin{equation}
  V_{\Gamma}^{O} = 2 \left( \begin{array}{cc}
      \frac{3}{2} \left( v_{33}^{{E}} + v_{44}^{{E}}\right) & 0 \\
      0 & \frac{3}{2} \left( v_{33}^{{E}} + v_{44}^{{E}} \right)
\end{array} \right),
\end{equation}
\begin{equation}
  U_{\Gamma} = 2 \left( \begin{array}{ccc}
      \frac{3}{2} \left( u_{11}^{{E}} + u_{22}^{{E}} \right) & 0 & 0\\
      0 & \frac{3}{2} \left( u_{11}^{{E}} + u_{22}^{{E}} \right) & 0 \\
      0 & 0 & 3\, u_{33}^{{E}} \end{array}\right),
\end{equation}
\begin{equation}
  S_{\Gamma}^{O} = \left( \begin{array}{ccc}
      - s_{11} & 0 & 0 \\ 0 & -s_{22} & 0 \\ 0 & 0 & s_{33}
\end{array} \right),
\end{equation}
and
\begin{equation}
  S_{\Gamma}^{E} = \left( \begin{array}{ccc}
      s_{11} & 0 & 0 \\ 0 & s_{22} & 0 \\ 0 & 0 & - s_{33}
\end{array} \right),
\end{equation}
where $t_{02}^{t} + 2\, t_{02}^{l,t} = t_{13}^{t} + 2\, t _{13}^{l,t}
= t_{43}^{t} + 2\, t_{43}^{l,t} = 0$ due to the SK decomposition. We
can break $H^{E}$ into three $2\times2$ diagonal blocks and $H^{O}$
into two $2\times2$ blocks and one $1\times1$ block. All blocks can
then be diagonalized analytically. Explicitly,

\begin{itemize}
\item $\alpha_{0}-\phi_{3}$ 
\begin{equation}
  \left( \begin{array}{cc} \Delta_{0} + 6\, v_{00}^{{E}} & \sqrt{2} \left(
        t_{03}^{t}
        + 2\, t_{03}^{l,t} \right) \\
      \sqrt{2} \left( t_{03}^{t} + 2\, t_{03}^{l,t} \right) & \Delta_{z} +
      6\, u_{33}^{{E}} - s_{33}
\end{array} \right)
\label{Gamma}
\end{equation}
\item $\alpha_{1}-\theta_{2}$ 
\begin{equation}
  \left( \begin{array}{cc}
      \Delta_{2} + 3 \left( v_{11}^{{E}} + v_{22}^{{E}} \right) & \sqrt{2} 
      \left( t_{12}^{t} + 2\, t_{12}^{l,t} \right) \\ \sqrt{2} \left( t_{12}^{t} 
        + 2\, t_{12}^{l,t} \right) & \Delta_{p} + 3\, \left( u_{11}^{{E}} 
        + u_{22}^{{E}} \right) + s_{22}
\end{array} \right)
\end{equation}
\item $\alpha_{2}-\theta_{1}$ 
\begin{equation}
  \left( \begin{array}{cc}
      \Delta_{2} + 3 \left( v_{11}^{{E}} + v_{22}^{{E}} \right) & \sqrt{2} 
      \left( t_{21}^{t} + 2\, t_{21}^{l,t} \right) \\ \sqrt{2} \left(t_{21}^{t}
        + 2\, t_{21}^{l,t} \right) & \Delta_{p} + 3\, \left( u_{11}^{{E}} 
        + u_{22}^{{E}} \right) + s_{11}
    \end{array} \right)
\end{equation}
\item $\alpha_{3}-\phi_{1}$ 
\begin{equation}
  \left( \begin{array}{cc}
      \Delta_{1} + 3 \left( v_{33}^{{E}} + v_{44}^{{E}} \right) & \sqrt{2} 
      \left( t_{31}^{t} + 2\, t_{31}^{l,t} \right) \\
      \sqrt{2} \left( t_{31}^{t} + 2\, t_{31}^{l,t} \right) & \Delta_{p} 
      + 3 \left( u_{11}^{{E}} + u_{22}^{{E}} \right) - s_{11}
\end{array} \right)
\end{equation}
\item $\alpha_{4}-\phi_{2}$
\begin{equation}
  \left(\begin{array}{cc}
      \Delta_{1} + 3 \left( v_{33}^{{E}} + v_{44}^{{E}} \right) & 
      \sqrt{2} \left( t_{42}^{t} + 2\, t_{42}^{l,t} \right) \\
      \sqrt{2} \left( t_{42}^{t} + 2\, t_{42}^{l,t} \right) & 
      \Delta_{p} + 3\, \left( u_{11}^{{E}} + u_{22}^{{E}} \right) - s_{22}
\end{array} \right)
\end{equation}
\item $\theta_{3}$ 
\begin{equation}
  \left( \begin{array}{c}
      \Delta_{z} + 6\, u_{33}^{{E}} + s_{33} \end{array} \right).
\end{equation}
\end{itemize}

The bands can be identified by matching their composition to the
results of the DFT-HSE06 calculations, (see Table
\ref{tab:orb_comp_G_DFT}). For instance, the valence band state at the
$\Gamma$ point is mostly composed by $d_{z^{2}}$ and $p_{z}$-orbitals.
Another band state (band number 6, as defined in Table
\ref{tab:orb_comp_G_DFT}) that has a similar orbital composition is
lower in energy than the valence band state. Therefore, we associate
the highest eigenvalue of (\ref{Gamma}) to the valence band energy,
while the lowest eigenvalue is put in correspondence with the band
number 6 energy. As a result, we find
\begin{eqnarray}
\hskip-1.0cm
  E_{\rm{valence}}(\Gamma) & = & \frac{\Delta_{0} + 6\, v_{00}^{{E}} 
    + \Delta_{z} + 6\, u_{33}^{{E}} - s_{33}}{2} \nonumber \\
  &  & + \sqrt{\frac{ \left( \Delta_{0} + 6\, v_{00}^{{E}} 
        - \Delta_{z} - 6\, u_{33}^{{E}} + s_{33} \right)^{2}}{4} 
    + 2\, \left( t_{03}^{t} + 2\, t_{03}^{l,t} \right)^{2}}
\end{eqnarray}
and
\begin{eqnarray}
\hskip-1.0cm
  E_{\rm{6}}(\Gamma) & = & \frac{\Delta_{0} + 6\, v_{00}^{{E}} 
    + \Delta_{z} + 6\, u_{33}^{{E}}-s_{33}}{2} \nonumber \\
  &  & - \sqrt{\frac{ \left( \Delta_{0} + 6\, v_{00}^{{E}} 
        - \Delta_{z} - 6\, u_{33}^{{E}} + s_{33} \right)^{2}}{4} 
    + 2\, \left( t_{03}^{t} + 2\, t_{03}^{l,t} \right)^{2}}.
\end{eqnarray}
Carrying out the same procedure for other blocks, we arrive at fully
analytical expressions for all bands at the $\Gamma$ point.

Thus each block correspond to a majority orbital composition and each
eigenvalue (two for each block $2\times 2$ $E_{-}$ and $E_{+}$) is
matched to a DFT value.

%%%%%%%%%%%%%%%%%%%%%%%%%%%%%%%%%%%%%
\subsection{$K$ point}

At the $K$ point, $k_{x}=2\pi/3a$ and $k_{y}=-2\pi/\sqrt{3}a$,
resulting in $z_{1} = e^{2i\pi/3}$, $z_{2} = -e^{i\pi/3}$, and $c_{1}
= c_{2} = c_{12} = - 1/2$. Then,
\begin{equation}
  T_{K}^{E,O} = \sqrt{2} \left( \begin{array}{ccc}
      i \sqrt{3}\, t_{01}^{l,t} & t_{02}^{t} - t_{02}^{l,t} & 0 \\
      i\sqrt{3}\, t_{11}^{l,t} & t_{12}^{t} - t_{12}^{l,t} & t_{13}^{t} 
      - t_{13}^{l,t} \\ t_{21}^{t} - t_{21}^{l,t} 
& i \sqrt{3}\, t_{22}^{l,t} & i \sqrt{3}\, t_{23}^{l,t}
\end{array} \right),
\end{equation}
\begin{equation}
  T_{K}^{O,E} = \sqrt{2} \left( \begin{array}{ccc}
      t_{31}^{t} - t_{31}^{l,t} & i \sqrt{3} t_{32}^{l,t} & 
      i \sqrt{3}\, t_{33}^{l,t} \\ i \sqrt{3}\, t_{41}^{l,t} & 
      t_{42}^{t} - t_{42}^{l,t} & t_{43}^{t} - t_{43}^{l,t}
\end{array} \right),
\end{equation}
\begin{equation}
  V_{K}^{E} = -  \frac{3}{2}\left( \begin{array}{ccc}
      2\, v_{00}^{E} & 0 & 0\\
      0 & v_{11}^{{E}} +  v_{22}^{{E}} & 0 \\
      0 & 0 &  v_{22}^{{E}} + v_{11}^{{E}}
\end{array} \right),
\end{equation}
\begin{equation}
  V_{K}^{O} = - \frac{3}{2}\left( \begin{array}{cc}
      v_{33}^{{E}} +  v_{44}^{{E}} & 0 \\
      0 &  v_{44}^{{E}} +   v_{33}^{{E}}
\end{array} \right),
\end{equation}
\begin{equation}
  U_{K}^{E,O} = - \frac{3}{2}\left( \begin{array}{ccc}
      u_{11}^{{E}} +  u_{22}^{{E}} & 0 & 0 \\
      0 &  u_{22}^{{E}} +  u_{11}^{{E}} & 0 \\
      0 & 0 & 2\, u_{33}^{{E}}
\end{array} \right),
\end{equation}
\begin{equation}
%\hskip-2.cm
  S_{K}^{O} = \left( \begin{array}{ccc}
-s_{11} & 0 & 0 \\ 0 & -s_{22} & 0 \\ 0 & 0 & s_{33}
\end{array} \right),
\quad
\end{equation}
and
\begin{equation}
  S_{K}^{E} = \left( \begin{array}{ccc}
      s_{11} & 0 & 0 \\ 0 & s_{22} & 0 \\ 0 & 0 & -s_{33}
\end{array} \right),
\end{equation}
where we have imposed $t_{03}^{t} = t_{03}^{t{L}}$ and $s_{11} =
s_{22}$.

The Hamiltonian matrix can be block diagonalized by a chiral
transformation \cite{Cap2013},
\begin{equation}
\alpha_{R2,L2} = \frac{\alpha_{1}\pm i\alpha_{2}}{\sqrt{2}},
\end{equation}
\begin{equation}
\alpha_{R1,L1} = \frac{\alpha_{3}\pm i\alpha_{4}}{\sqrt{2}},
\end{equation}
\begin{equation}
\theta_{R,L} = \frac{\theta_{1}\pm i\theta_{2}}{\sqrt{2}},
\end{equation}
\begin{equation}
\phi_{R,L} = \frac{\phi_{1}\pm i\phi_{2}}{\sqrt{2}}.
\end{equation}
The other variables, $\alpha_{0}$, $\theta_{3}$, and $\phi_{3}$,
remain the same. Thus, we have the new state vector
\begin{eqnarray}
  \psi^{T} = \left(\alpha_{0}, \alpha_{L2}, \alpha_{R2}, \theta_{L}, \theta_{R}, 
      \phi_{3}, \alpha_{L1}, \alpha_{R1}, \phi_{L}, \phi_{R},  \theta_{3}  \right).
\end{eqnarray}
The result is 
\begin{equation}
H = \left( \begin{array}{cc} A & 0 \\ 0 & B \end{array} \right),
\end{equation}
where
\begin{eqnarray}
\hskip-2.0cm
  A  = \left( \begin{array}{cccc}
      \Delta_{0} + V_{00}^{E} & 0 & 0 & \\
      0 & \Delta_{2} + V_{11}^{E} & 0 & \\
      0 & 0 & \Delta_{2} + V_{22}^{E}  & \cdots \\
      0 & -\frac{i}{\sqrt{2}} (K_{11}-K_{21}+K_{12}+K_{22}) & 0 & \\
      -i(K_{01}-K_{02}) & 0 & 0 & \\
      0 & 0 & K_{13}-K_{23} & \end{array} \right. \\ 
\hskip-1.0cm\left.
\begin{array}{cccc}
      & 0 & i(K_{01}-K_{02}) & 0\\
      & \frac{i}{\sqrt{2}} (K_{11}-K_{21}+K_{12}+K_{22}) & 0 & 0\\
      \cdots & 0 & 0 & K_{13}-K_{23}\\
      & \Delta_{p} + U_{11} + s_{11} & 0 & 0 \\
      & 0 & \Delta_{p} + U_{22} + s_{22} & 0 \\
      & 0 & 0 & \Delta_{z} + U_{33} - s_{33}
\end{array} \right)
\nonumber
%\end{split}
\label{eq:matrixA}
\end{eqnarray}
and
\begin{eqnarray}
\hskip-2.0cm
B = \left( \begin{array}{ccc}
\Delta_{1} + V_{33}^{O} & 0 & \\
0 & \Delta_{1} + V_{44}^{O} & \\
0 & \frac{1}{\sqrt{2}} (K_{31}-K_{41}-K_{32}-K_{42}) & \cdots \\
0 & 0 & \\
-i(K_{33}-K_{43}) & 0 &
\end{array} \right. \\ \left.
\begin{array}{cccc}
 & 0 & 0 & i(K_{33}-K_{43}) \\
 & \frac{1}{\sqrt{2}} (K_{31}-K_{41}-K_{32}-K_{42}) & 0 & 0 \\
\cdots & \Delta_{p} + U_{11} - s_{11} & 0 & 0 \\
 & 0 & \Delta_{p} + U_{22} - s_{11} & 0 \\
 & 0 & 0 & \Delta_{z} + U_{33} + s_{33}
\end{array} \right)
\nonumber
\label{eq:matrixB}
\end{eqnarray}
The following elements have been introduced: 
\begin{eqnarray}
K_{01}  & = & \sqrt{3}t_{01}^{l,t}, \qquad
K_{02}     =      t_{02}^{t}-t_{02}^{l,t},
\nonumber\\
K_{11} &  =  & \sqrt{3}t_{11}^{l,t}, \qquad
K_{12}     =       t_{12}^{t}-t_{12}^{l,t}, \quad\;\;\;
K_{13}     =       t_{13}^{t}-t_{13}^{l,t}, 
\nonumber\\
K_{21} & = & t_{21}^{t}-t_{21}^{l,t},\quad\;\;\;
K_{22}     =     \sqrt{3}t_{22}^{l,t},\qquad
K_{23}     =      \sqrt{3}t_{23}^{l,t},
\nonumber\\
K_{31} & = & t_{31}^{t}-t_{31}^{l,t},\quad\;\;\;
K_{32}     =     \sqrt{3}t_{32}^{l,t},\qquad
K_{33}     =     \sqrt{3}t_{33}^{l,t},
\nonumber\\
K_{41} & = & \sqrt{3}t_{41}^{l,t}, \qquad
K_{42}     =     t_{42}^{t}-t_{42}^{l,t},\quad\;\;\;
K_{43}     =    t_{43}^{t}-t_{43}^{l,t}.
\nonumber
\end{eqnarray}
Using the SK decomposition, one finds that several combinations of
these coefficients yield zero. These simplications have already been
implemented in Eqs. (\ref{eq:matrixA}) and (\ref{eq:matrixB}). The $H$
matrix breaks up into five $2\times2$ blocks and one $1\times 1$
block,
\begin{itemize}
\item
$\alpha_{0}-\theta_{R}$: 
\begin{equation}
  \left( \begin{array}{cc}
      \Delta_{0} + V_{00}^{E} & i(K_{01}-K_{02}) \\
      -i (K_{01}-K_{02}) & \Delta_{p} + U_{11}^{E} + s_{11}
\end{array} \right)
\end{equation}
\item $\alpha_{L2}-\theta_{L}$: 
\begin{equation}
  \left(\begin{array}{cc}
      \Delta_{2} + V_{11}^{E} & i (K_{11}-K_{21}+K_{12}+K_{22})/\sqrt{2} \\
      - i (K_{11}-K_{21}+K_{12}+K_{22})/\sqrt{2} & \Delta_{p} + U_{22}^{E} + s_{11}
\end{array} \right)
\end{equation}
\item $\alpha_{R2}-\phi_{3}$: 
\begin{equation}
  \left( \begin{array}{cc}
      \Delta_{2} + V_{22}^{E} &  K_{13}-K_{23} \\
      K_{13}-K_{23} & \Delta_{z} + U_{33}^{E} - s_{33}
\end{array} \right)
\end{equation}
\item $\alpha_{R1}-\phi_{L}$: 
\begin{equation}
  \left( \begin{array}{cc}
      \Delta_{1} + V_{44}^{O} & (K_{13}-K_{41}-K_{23}-K_{42})/\sqrt{2} \\
      (K_{13}-K_{41}-K_{23}-K_{42})/\sqrt{2} & \Delta_{p} + U_{22}^{O} - s_{11}
\end{array} \right)
\end{equation}
\item $\alpha_{L1}-\theta_{3}$: 
\begin{equation}
  \left( \begin{array}{cc}
      \Delta_{1} + V_{33}^{O} & i (K_{33}-K_{43}) \\
      -i (K_{33}-K_{43}) & \Delta_{z} + U_{33}^{O} + s_{33}
\end{array} \right)
\end{equation}
\item $\phi_{R}$:
\begin{equation}
\Delta_{p} + U_{11}^{O} - s_{11}
\end{equation}
\end{itemize}

According to the DFT-HSE06 calculations, at the $K$ point, the
conductance band is mainly composed by $d_{3z^{2}-r^{2}}$, $p_{x}$,
and $p_{y}$ orbitals, while the valence band is mainly formed by
$d_{x^{2}-y^{2}}$, $d_{xy}$, $p_{x}$, and $p_{y}$ orbitals. Therefore,
the conductance band can be obtained from $\alpha_{0}-\theta_{L}$
variables, while the valence band comes from the
$\alpha_{L2}-\theta_{R}$ combinations, resulting in the expressions
\begin{eqnarray}
\hskip-1.5cm
  E_{\rm{conductance}}(K) & = & \frac{\Delta_{0} + V_{00}^{E} 
    + \Delta_{p} + U_{11}^{E}+s_{11}}{2} \nonumber \\ & & 
  + \sqrt{\frac{(\Delta_{0} + V_{00}^{E} - \Delta_{p} - U_{11}^{E} 
      - s_{11})^{2}}{4} + (K_{01}-K_{02})^{2}}
\end{eqnarray}
and
\begin{eqnarray}
\hskip-1.5cm
  E_{\rm{valence}}(K) & = & \frac{\Delta_{2} + V_{11}^{E} + \Delta_{p} 
    + U_{22}^{E}+s_{11}}{2} \nonumber \\ & & + \sqrt{\frac{(\Delta_{2} 
      + V_{11}^{E} - \Delta_{p} - U_{22}^{E} - s_{11})^{2}}{4} 
    + \frac{(K_{11} - K_{21} - K_{12} - K_{22})^{2}}{2}}.
\end{eqnarray}

Following the same procedure for other blocks, we find analytical
expressions for nearly all energies at the $K$ point. Thus, each block
corresponds to a majority orbital composition and each eigenvalue (two
for each block $2\times 2$) is matched to a DFT-HSE06 value.

%%%%%%%%%%%%%%%%%%%%%%%%%%%%%%%%%%%%%%%%%%%%%
\section{Alternative implementation of unsymmetrized band
  equations and comparison with the tight-binding model of Cappelluti
  {\it et al}.}
\label{subsec:Appendix D}

The essential difference between our construction of the 11-orbital
and that of Cappelluti {\it et al} \cite{Cap2013,Cap2014} comes from
their inclusion of two phase factors in the Bloch state equation,
namely,
\begin{eqnarray}
  \hskip-1.cm
  |{\bf k} \rangle = \sum_{{\bf r}_i}
  e^{i{\bf k} \cdot {\bf r}_i} && \left[ \sum_{\mu=0}^4
    \alpha_{{\bf k},\mu} |  {\bf r}_i;
    d_{\mu} \rangle + \sum_{\mu=1}^3
    (\beta_{{\bf k},\mu} e^{i {\bf k} \cdot {\bf \delta}_{1-}}|
    {\bf r}_i + {\bf \delta}_{1-} ; p_{\mu} \rangle \right.
    \nonumber\\ 
&& \left.   + \tau_{{\bf k},\mu} e^{i{\bf k} \cdot {\bf \delta}_{1+}}
    | {\bf r}_i + {\bf \delta}_{1+} ; p_{\mu} \rangle
    ) \right].
\end{eqnarray}
Comparing this equation with Eq. (\ref{eq:Bloch}), we notice the extra
phase factors in the amplitudes of the S atomic orbitals. While the
phase factors have no impact on the eigenvalue secular equation, they
do change the matrices containing hopping amplitudes between Mo and S
atoms. For instance, our $T^t$ matrix of Eq. (\ref{eq:Tt}) would
change to
\begin{equation}
  T^t = \left[ t^t + z_2^{\ast}  \left( t^{l,t} + z_1 t^{r,t}
    \right) \right] e^{{ik} \cdot \delta_{1+}}.
\end{equation}
A second yet important difference between their work and ours is on
the notation and organization of the hopping matrices.

To facilitate a direct comparison between our model and that of
Refs. \cite{Cap2013,Cap2014}, we begin by swapping the second and
third rows and corresponding columns in Eq. (\ref{eq:eigen_matrix}),
\begin{equation}
  \left(\begin{array}{ccc}
      h^S + U & ^{} (T^t)^{\dagger} & s\\
      T^t & h^{\rm Mo} + V & T^b\\
      s & (T^b)^{\dagger} & h^{{S}} + U
  \end{array}\right) \left(\begin{array}{c}
    \tau\\
    \alpha\\
    \beta
  \end{array}\right) = E \left(\begin{array}{c}
    \tau\\
    \alpha\\
    \beta
  \end{array}\right).
\end{equation}
Next we introduce their auxiliary quantities $\xi = k_x a/2$ and $\eta
= \sqrt{3}\, k_y a/2$, which allows us to rewrite the coefficients in
Eqs. (\ref{eq:c1}), (\ref{eq:c2}), and (\ref{eq:c12}) as
\begin{equation}
  c_1 = \cos (2 \xi),
\end{equation}
\begin{equation}
  c_2 = \cos (\xi + \eta),
\end{equation}
and
\begin{equation}
  c_{12} = \cos (\xi - \eta).
\end{equation}
Also, $z_1 = e^{2i\xi}$, $ z_2 = e^{^{i (\xi + \eta)}}$, and
\begin{eqnarray}
  e^{i{\bf k} \cdot {\bf \delta}_1} & = & e^{2i \eta/3} \\
  e^{i{\bf k} \cdot {\bf \delta}_2} & = & e^{- i \left( \xi + \eta/3 \right)} \\
  e^{i{\bf k} \cdot {\bf \delta}_3} & = & e^{i \left( \xi - \eta/3 \right)}.
\end{eqnarray}

The correspondence between our block matrices and theirs is the
following (phase factors set to zero in the appropriate hopping
amplitudes):
\begin{itemize}
\item $h^S + U \leftrightarrow H_{{pt}, {pt}} = H_{{pb}, {pb}}$ with
\begin{equation}
  H_{{pt}, {pt}} =
  \left(\begin{array}{ccc}
      H_{{xx}} & H_{{xy}} & 0\\
      H_{{xy}}^\ast  & H_{{yy}} & 0\\
      0 & 0 & H_{{zz},z}
    \end{array} \right)
\end{equation}
\item $S \leftrightarrow H_{{pt}, {pb}}$, with
\begin{equation}
  H_{{pt}, {pb}} = 
  \left(\begin{array}{ccc}
      V_{pp\pi} & 0 & 0\\
      0  & V_{pp\pi} & 0\\
      0 & 0 & V_{pp\sigma}
\end{array}\right)
\end{equation}
\item $h^{{Mo}} + V \leftrightarrow H_{d, d}$, with
\begin{equation}
  H_{d, d} =
  \left(\begin{array}{ccccc}
      H_{z^{2}z^{2}} & H_{z^{2}x^{2}} &
      H_{z^{2}xy}  & 0 & 0 \\
      H_{z^{2}x^{2}}^\ast & H_{x^{2}x^{2}} &
      H_{x^{2}xy}  & 0 & 0 \\
      H_{z^{2}xy}^\ast & H_{x^{2}xy}^\ast & H_{xy,xy}  &
      0 & 0\\
      0 & 0 & 0 & H_{xz,xz} & H_{xz,yz} \\
      0 & 0 & 0 & H_{xz,yz}^\ast & H_{yz,yz} 
\end{array}\right)
\end{equation}
\item $T^t \leftrightarrow H_{d, {pt}}$, with
\begin{equation}
H_{d, {pt}} = ^{} \left(\begin{array}{ccc}
  H_{z^2 x} & H_{z^2 y} & H_{z^2 z}\\
  H_{x^2 x}  & H_{x^2 y} & H_{x^2 z}\\
  H_{{xy}, x} & H_{{xy}, y} & H_{{xy}, z}\\
  H_{{xz}, z} & H_{{xz}, y} & H_{{xz}, z}\\
  H_{{yz}, x} & H_{{yz}, y} & H_{{yz}, z}\\
  &  & 
\end{array} \right)
\end{equation}
\item $T^b \leftrightarrow H_{d, {pb}}$, with
\begin{equation}
H_{d, {pb}} = ^{} \left(\begin{array}{ccc}
  H_{z^2 x} & H_{z^2 y} & -H_{z^2 z}\\
  H_{x^2 x}  & H_{x^2 y} & -H_{x^2 z}\\
  H_{{xy}, x} & H_{{xy}, y} & -H_{{xy}, z}\\
  -H_{{xz}, z} & -H_{{xz}, y} & H_{{xz}, z}\\
  -H_{{yz}, x} & -H_{{yz}, y} & H_{{yz}, z}\\
  &  & 
\end{array} \right).
\end{equation}
\end{itemize}

All matrix elements are identical to those of Ref. \cite{Cap2013},
except for the matrices $H_{d,pt}$ and $H_{d,pb}$, which are explicitly
defined below:
\begin{eqnarray}
H_{z^2 x} & = & - 2\, \sqrt{3}\, E_1\, \sin (\xi)\, d_1 \\
H_{z^2 y} & = & 2\, E_1\, C_2 \\
H_{z^2 z} & = & E_2\, C_1 \\
H_{x^2 x} & = & - 2 \sqrt{3}\, \left( \frac{1}{3}\, E_5 - E_3 \right)\, \sin (\xi)\, d_1\\
H_{{xy}, y} & = & H_{x^2 x} \\
H_{{xz}, z} & = & - 2\, \sqrt{3}\, E_8\, \sin (\xi)\, d_1 \\
H_{{yz}, x} & = & H_{{xz}, y} \\
H_{{yz}, z} & = & 2\, E_8\, C_2, \\
  H_{x^2 y} & = & - 2\, E_3\, C_3 + 2i\, E_5\, \cos (\xi)\, d_1 \\
  H_{x^2 z} & = & 2\, E_4\, C_2 \\
  H_{{xy}, x} & = & - \frac{2}{3}\, E_5\, C_3 + 6i\, E_3\, \cos (\xi)\, d_1 \\
  H_{{xy}, z^{}} & = & - 2 \sqrt{3}\, E_4\, \sin (\xi) d_1 \\
  H_{{xz}, y} & = & - 2 \sqrt{3} \left( \frac{1}{3}\, E_6 - E_7 \right) \sin
  (\xi)\, d_1 \\
  H_{{xz}, x} & = & - \frac{2}{3}\, E_6\, C_3 + 6i\, E_7 \cos (\xi)\, d_1 \\
  H_{{yz}, y} & = & - 2\, E_7\, C_3 + 2i\, E_6\, \cos (\xi)\, d_1.
\end{eqnarray}
In all these equations the quantities $l_1$, $l_2$ and $l_3$, as well
$E_i$, $i=1,\ldots,8$, follow the definitions of Ref. \cite{Cap2013};
for instance,
\begin{eqnarray}
l_1 & = & \cos (2\xi) + 2\, \cos (\xi)\, \cos (\eta) \\
l_2 & = & \cos (2\xi)  - \cos (\xi)\, \cos (\eta) \\
l_3 & = & 2\, \cos (2\xi) +\cos (\xi)\, \cos (\eta).
\end{eqnarray}
The coefficients $C_1$, $C_2$, $C_3$, and $d_1$ become more compact
without the inclusion of phases, namely,
\begin{eqnarray}
C_1 & = & - 1 - 2\, \cos (\xi)\, \cos (\eta) + 2 i\,\cos (\xi)\, \sin (\eta) \\
C_2 & = & - 1 + \cos (\xi)\, \cos (\eta) - i\cos (\xi)\, \sin (\eta) \\
C_3 & = & \cos (\xi)\, \cos (\eta) - i\cos (\xi)\, \sin (\eta) + 2 \\
d_1 & = & i\cos (\eta) + \sin (\eta).
\end{eqnarray}
%

%%%%%%%%%%%%%%%%%%%%%%%%%%%%%%%%%%%%%%
% References
\section*{References}
%%%%%%%%%%%%%%%%%%%%%%%%%%%%%%%%%%%%%%

%%%%%%%%%%%%%%%%%%%%%%%%%%%%%%%%%%%%%

\end{document}